%% file: main.tex
\documentclass[letterpaper,twocolumn,10pt]{article}
\usepackage{usenix-2020-09}

\usepackage{amsmath}
\usepackage{amssymb}
\usepackage{booktabs}
\usepackage{array}
\usepackage{tabularx}
\usepackage{longtable}
\usepackage{ragged2e}
\usepackage{capt-of}
\usepackage{pdflscape}
\usepackage{CJKutf8}
\usepackage{graphicx}
\usepackage[caption=false,font=footnotesize]{subfig}
\AddToHook{env/figure/begin}{%
  \setlength{\abovecaptionskip}{4pt}%
  \setlength{\belowcaptionskip}{0pt}}
\AddToHook{env/figure*/begin}{%
  \setlength{\abovecaptionskip}{4pt}%
  \setlength{\belowcaptionskip}{0pt}}
\usepackage{multirow}
\usepackage{xcolor}
\usepackage{listings}
\usepackage{enumitem}
\usepackage[left,mathlines]{lineno}
\newif\ifdraftlinenumbers
\draftlinenumbersfalse

\usepackage{url}
\usepackage{xspace}
\usepackage{tikz}
\usetikzlibrary{arrows.meta,positioning}
\input{figs/dp-agent-icon}

\newcommand{\figref}[1]{\mbox{Figure~\ref{#1}}}
\newcommand{\tabref}[1]{\mbox{Table~\ref{#1}}}
\makeatletter
\newcommand{\figurepanellabel}[1]{%
  \begingroup
  \advance\c@figure\@ne
  \refstepcounter{subfigure}%
  \label{#1}%
  \endgroup}
\makeatother
\newcommand{\sys}{Herschel\xspace}

\newcommand{\para}[1]{%
  \par\addvspace{3pt}%
  \noindent\textbf{#1}\hspace{0.5em}\ignorespaces
}
\definecolor{ourrevisioncolor}{HTML}{000000}
\newenvironment{ourrevision}{\begingroup\color{ourrevisioncolor}}{\par\endgroup}

\begin{document}
% Keep paragraph spacing natural; absorb residual page space at the bottom.
\raggedbottom
% Avoid a hyphenated word continuing at the top of the next column/page.
\brokenpenalty=10000

\title{\sys: Continuous Optimization of Production LLM Inference through On-Demand Profiling}
% \title{\sys: From Fleet-Wide Profiling to Continuous Optimization of LLM Inference}
% \title{\sys: Fleet-Wide Profiling for Continuous Optimization of LLM Inference Services}
% Public arXiv version; author order and affiliations supplied by the authors.
% \vspace{-.3in}
\author{
{\normalfont Luping Wang\thanks{\raggedright Corresponding author: Luping Wang, \texttt{chamu.wlp@alibaba-inc.com}.},\quad Weigao Chen,\quad Yifei Wu,\quad Yonghe Zhang,\quad Rui Zhang}\\[2pt]
{\normalfont Wenchao Wu,\quad Jiyu Luo,\quad Haoran Geng,\quad Xin Yang,\quad Chen Cao}\\[2pt]
{\normalfont Yuemin Wu,\quad Cheng Huang,\quad Guodong Yang,\quad Liping Zhang}\\[2pt]
{\normalfont\normalsize Alibaba Group}
}
\maketitle
\ifdraftlinenumbers\linenumbers\fi
\input{section/abstract}

\input{section/intro}

\input{section/background}

\input{section/motivation}

\input{section/overview}

\input{section/aprof}

\input{section/analyzer}

\input{section/optimizer}
\input{figs/evaluation-overhead-figure}

\input{section/eval}

\input{section/case-studies}

\input{section/discussion}

\input{section/related_work}

\input{section/conclusion}

% NSDI '27 double-blind submission: acknowledgments must be omitted.
% \input{section/acknowledgement}
\bibliographystyle{plain}
\bibliography{references,analyzer-patterns}

% \onecolumn
% \appendix
% \input{section/appendix-observations}

\end{document}

%% file: figs/dp-agent-icon.tex
\colorlet{DPagentink}{black!55}
\DeclareRobustCommand{\DPAgentIcon}{%
\begin{tikzpicture}[x=1pt,y=-1pt,scale=0.82,baseline=-1.7pt,
 line cap=round,line join=round]
\path[use as bounding box] (-7,-7) rectangle (7,6.5);
\draw[DPagentink,line width=.58pt] (0,-4)--(0,-6);
\fill[DPagentink] (0,-6.1) circle[radius=.62pt];
\path[draw=DPagentink,fill=white,line width=.58pt,rounded corners=.35pt]
 (-6.6,-1.5) rectangle (-5.1,1.8)
 (5.1,-1.5) rectangle (6.6,1.8);
\path[draw=DPagentink,fill=white,line width=.58pt,rounded corners=.8pt]
 (-5.2,-4) rectangle (5.2,5);
\fill[DPagentink] (-3.0,-2.0) rectangle (-1.8,-.8)
 (1.8,-2.0) rectangle (3.0,-.8);
\draw[DPagentink,line width=.55pt]
 (-2.3,1.0) rectangle (2.3,3.4);
\end{tikzpicture}}

%% file: section/abstract.tex
\begin{abstract}
Model-as-a-service platforms call for continuous optimization as complex
serving conditions expose inefficiencies missed before deployment.
Detailed always-on profiling can incur substantial overhead, while
lightweight collection omits information needed for diagnosis.
We present \sys, a continuous optimization system for production
large language model (LLM) inference. Our key insight is that \emph{adaptive,
on-demand profiling can provide rich full-stack evidence without
continuous collection}. 
\sys safely attaches to and detaches from selected running processes
without engine changes or restarts, and adapts coverage as
investigations reveal missing evidence.
\sys reconstructs operator
executions and cross-process dependencies to identify inefficiency
mechanisms and suggest solutions using applicable reference fixes.
AI agents implement and test engine
and kernel changes under controlled conditions that preserve the
triggering workload and dependencies, with expert review before
deployment. 
Controlled tests show active-collection overhead below 0.5\% for
time to first token and 7\% for time per output token.
Bounded windows, typically 30\,s, avoid the continuous cost of
always-on tracing.
Over six months, \sys collected $\sim$17,000 traces across
over 120 model variants and more than 10 accelerator types, identifying
inefficiency patterns in 23\% of the traces. Representative findings guide
widely deployed optimizations, including restructured synchronization,
removal of unused computation, and improved operator implementations.
\end{abstract}

%% file: section/intro.tex
\section{Introduction}
\label{sec:intro}

\begin{figure}[t]
  \centering
  % Figure-local spacing: trim panel whitespace and tighten both caption gaps.
  \captionsetup[subfloat]{captionskip=2pt,farskip=0pt}
  \setlength{\abovecaptionskip}{4pt}
  \setlength{\belowcaptionskip}{0pt}
  \subfloat[Flagship model performance: 8 weeks of post-release optimization.]{%
    \includegraphics[width=.49\columnwidth,trim=0 6pt 0 0,clip]{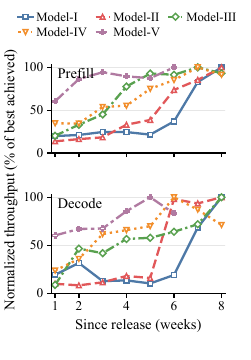}%
    \label{fig:motivation-ramp}}%
  \hfill
  \subfloat[Complexity in serving configurations and workloads.]{%
    \includegraphics[width=.49\columnwidth,trim=0 6pt 0 0,clip]{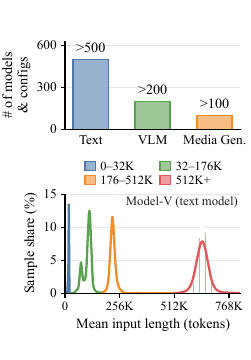}%
    \label{fig:motivation-serving-conditions}}
  \caption{Continuous optimization under complex serving conditions on MaaS platforms.}
  \label{fig:motivation-overview}
  \vspace{-3pt}
\end{figure}

Model-as-a-service (MaaS) platforms pursue \emph{day-one availability},
offering new models on their release day~\cite{meta-llama31-launch,anthropic-claude37-launch}.
Pre-deployment tuning covers selected workloads and can leave performance
inefficiencies unaddressed.
\figref{fig:motivation-ramp}
shows performance improvements across five of our flagship models over up to
eight weeks after release.
These post-release gains are achieved through repeated cycles of analysis,
improvement, and validation---\emph{continuous optimization}.

Continuous optimization is challenging under complex \emph{serving conditions}.
MaaS spans many combinations of models and deployment configurations,
each with varying workloads.
\figref{fig:motivation-serving-conditions}
illustrates this diversity.
Locating avoidable waits, missed communication--computation overlap,
redundant state preparation, and unused computation can require examining
dependencies among engine components.
Operator efficiency also depends on matching data layouts and parallelism
to input shapes and target GPU architectures.
Moreover, the same inefficiency can be prominent in one workload yet negligible
in another.

Identifying these inefficiencies and assessing their significance requires full-stack
online profiling, which remains challenging with existing approaches.
Anomaly-driven diagnosis may overlook persistent inefficiencies that do
not deviate from historical baselines~\cite{striatrace}.
For always-on profiling, retaining rich full-stack detail conflicts with
limiting sustained collection overhead.
Detailed tracing in PyTorch Profiler and Nsight Systems can incur
substantial overhead~\cite{pytorchprofiler,nsight-systems},
while lightweight collection restricts the calls and runtime state
available for investigation.
A cloud provider must also decouple collection from engine releases:
remote triggering alone can still require launch-time
support~\cite{dynolog}.

To address these constraints, our first insight is that
\emph{full-stack profiling can be adaptive without being always on.}
Runtime attachment enables on-demand profiling of running services without
engine changes or restarts.
Detailed recording is limited to bounded windows, retaining rich
full-stack evidence while avoiding the continuous cost of always-on tracing.
Collection need not wait for anomalies; recording coverage can be adjusted in later windows to fill evidence gaps.

Furthermore, turning profiling observations into deployable optimizations still requires substantial expert
effort, both to pinpoint root causes across different serving conditions and to develop effective fixes.
Our second insight is that \emph{inefficiency mechanisms and optimization
strategies recur across serving conditions.}
These mechanisms guide root-cause identification, while reference fixes
give AI agents a starting point for implementation and testing.
Current evidence helps assess applicability and identify correctness
constraints to validate.

We present \textbf{\sys}, a system for continuous optimization of production
LLM inference with three components:

\begin{itemize}[itemsep=3pt,parsep=0pt,topsep=3pt,partopsep=0pt]
  \item \textbf{Hprof} uses runtime-attached probes
    for on-demand profiling within bounded windows. On supported runtimes,
    it captures Python calls and supported object fields, accelerator API
    calls, and device activity from running services without application
    source-code changes or restarts. A fleet control plane selects processes and
    recording windows for serving conditions; investigations can adapt
    coverage as evidence needs change
    (\S\ref{sec:wprof}).

  \item \textbf{Analyzer} then reconstructs operator
    executions and cross-process dependencies from these records and checks
    the conditions of recurring engine and operator inefficiency patterns.
    It uses confirmed mechanisms and
    applicable reference fixes to produce optimization opportunities with
    supporting evidence and suggested solutions. When evidence is
    insufficient to confirm a mechanism, Analyzer requests expert-approved
    supplementary profiling (\S\ref{sec:analyzer}).

  \item \textbf{Optimizer} uses these opportunities
    to guide engine and kernel agents in implementing and validating
    candidate changes. Experiments use controlled deployment
    scales while retaining the workload conditions and dependencies that
    expose the problem. Correctness and performance feedback guide
    successive revisions; expert review determines which changes proceed
    to canary validation and staged rollout (\S\ref{sec:optimizer}).
\end{itemize}

We evaluate \sys in controlled tests, where Hprof's active-collection
overhead stays below 0.5\% for TTFT and below 7\% for TPOT across the
tested model sizes and workloads.
Bounded windows, typically 30\,s, avoid the continuous cost of
always-on tracing, with safe attachment and detachment.
Our six-month production study spans over 120 model variants and more
than 10 accelerator types, identifying inefficiency patterns in 23\%
of approximately 17,000 traces.
Findings include unused computation, repeated state preparation, and
operator inefficiencies at small input shapes.
These findings guided optimizations that are now widely deployed
after expert review and canary validation.

%% file: section/background.tex
\section{Background}
\label{sec:background}

\para{Production MaaS inference.}
Model-as-a-service (MaaS) platforms serve language, reasoning,
multimodal, and generative models through
APIs~\cite{alibaba-model-studio,model-garden}.
MaaS platforms support diverse workloads with different performance requirements.
Chat requires responsive replies; coding assistants process
repository context; agents interleave model calls with sandboxed
tool execution~\cite{anthropic-managed-agents}.
Repeated calls can accumulate long, shared context, increasing
input-processing and storage demands, while dependencies between calls make task completion time sensitive to
generation latency~\cite{semianalysis-agentx}.
\emph{Fast modes} prioritize per-request output tokens per second
(TPS)~\cite{alibaba-fast-mode,claude-fast-mode}, whereas asynchronous
batch services trade higher latency for lower
prices~\cite{claude-batch-processing}.
Platforms also offer confidential inference for workloads requiring
code and data protection~\cite{nvidia-confidential-ai}.

\para{Inference workflow.}
Inference engines schedule requests and manage their
state~\cite{vllm,sglang,tensorrtllm}. In autoregressive generation,
\emph{prefill} processes prompts to produce the first output token.
Standard attention layers also populate a key--value (KV) cache, retaining
states for reuse. During \emph{decode}, each step uses the latest token and
cached states to generate the next token and extend the cache, until an end
token or length limit is reached. Requests execute in batches, whose composition
and computational load depend on arrivals~\cite{servegen}.

\para{Maximizing inference efficiency.}
LLM service-level objectives (SLOs) cover latency and throughput.
Latency requirements bound
\emph{time to first token} (TTFT), from request arrival to first output, and
\emph{time per output token} (TPOT), the mean interval between consecutive
output tokens~\cite{distserve}. Throughput requirements specify the minimum
number of tokens processed per second across requests~\cite{llmpilot-slo-2026}.

MaaS providers seek to maximize token throughput per GPU under
latency SLOs to improve serving cost-efficiency.
They combine greater compute capability with advanced hardware
architectures, such as rack-scale systems~\cite{nvidia-rubin-pod}.
\emph{Engine optimizations} include KV prefix reuse through
distributed cache architectures~\cite{mooncake,lmcache} and
speculative decoding~\cite{speculative-decoding}, while
\emph{operator optimizations} target specific hardware and input
shapes (tensor dimension sizes)~\cite{flashinfer}.
\emph{Serving architectures} separate prefill from decode
(PD disaggregation), decouple attention from feed-forward computation
(AFD), or share prefill capacity across clusters, enabling independent
resource provisioning and flexible
placement~\cite{distserve,megascale-infer,prefill-as-a-service}.
% Together, these efforts seek higher throughput and faster responses.

%% file: section/motivation.tex
\section{Motivation}
\label{sec:motivation}

% \subsection{Continuous Optimization in MaaS}
% \label{sec:motivation-continuous}
% % Preserve downstream callbacks to the motivating production case.
% \label{sec:motivation-online}

Effective optimization after deployment requires understanding how
execution and serving conditions affect performance.

\input{figs/motivation-ab-figure}

\para{Production exposes cross-component inefficiencies.}
Isolated tuning can overlook engine coordination costs, redundant state
preparation, and unused computation.
In a GLM-5.2~\cite{glm52,glm52-blog} deployment, an upstream KV-service
operation delays command dispatch, leaving Workers waiting and GPUs idle,
as shown in \figref{fig:eval-kvcache-evidence}.
Such dependencies also govern communication--computation overlap, state
reuse, and dataflow across operator kernels.

\para{Optimization priorities depend on serving conditions.}
Separately, outer KV preparation occupied over 8\% of a GLM-5.2 Worker's
active window with high KV reuse and short executions; another service
with fewer hits spent under 1\% of pooled capture time on this work.
Traffic also determines frequent operator shapes (tensor dimensions) and
request lengths. An operator with fast compute kernels can still incur
costly preparation or result combination; layouts suited to one shape
or device may be inefficient for another.
In one deployment, the most frequent Gated DeltaNet
query-token/context-length bin reached
\emph{only} 14.7\% of the compute/bandwidth SOL bound~\cite{solar2026,sol-execbench},
as shown in \figref{fig:motivation-shape-efficiency}.

\para{The need for online profiling.}
% Identifying optimization opportunities requires online execution evidence
% tied to LLM serving conditions, including dependencies and workloads
% that isolated offline tests may miss.
Identifying optimization opportunities requires execution evidence tied to LLM
serving conditions. Online profiling captures
dependencies and workloads that isolated offline tests may miss.

\subsection{Existing Work and Challenges}
\label{sec:motivation-challenges}
\label{sec:motivation-optimization}

% Continuous optimization targets sustained
% performance degradation and persistent inefficiency, including
% implementations that have never performed better. Its objective is to
% establish and realize improvement opportunities, rather than stop at
% detecting or explaining anomalously slow requests.

Existing profiling and diagnosis approaches address parts of continuous
optimization; obtaining sufficient evidence and turning findings into
improvements still pose four challenges.

\para{Anomaly-driven diagnosis.}
Anomaly detectors identify deviations from expected behavior, so a
persistent inefficiency may remain within their baselines.
StriaTrace combines semantic spans and GPU activity for online diagnosis
and also supports manual investigation of persistent degradation~\cite{striatrace}.
Our concern is obtaining the evidence needed for an optimization
investigation when relevant calls, object fields, or dependencies fall
outside existing instrumentation.

\par\addvspace{4pt}
\noindent\emph{Challenge 1: Identify sustained performance
inefficiencies without relying on outlier detection or being
limited by predefined instrumentation coverage.}
\par\addvspace{4pt}

\para{Always-on profiling struggles to provide rich evidence at low overhead.}
Continuous collection must balance execution detail against online cost.
PyTorch Profiler and Nsight Systems capture detailed traces, including
CUDA activity through CUPTI~\cite{nvidia-cupti}, but collecting these \emph{continuously} can
impose substantial overhead~\cite{pytorchprofiler,kineto,nsight-systems}.
Lightweight approaches such as StriaTrace~\cite{striatrace} reduce overhead through sampled stacks and selective
semantic spans alongside GPU activity, omitting CUDA Runtime and
Driver API records.
Restricting collection further to regular py-spy stack samples leaves
out invocation-specific object state and GPU execution~\cite{pyspy}.
Such restricted coverage can leave gaps in optimization evidence:
assessing operator efficiency requires shapes, constituent kernels, and full
invocation duration; dependency and overlap analysis requires
cross-process relationships
(\S\ref{sec:analyzer-evidence}).

% \noindent\emph{Challenge 2: Enable precise online localization of
% performance inefficiencies with both sufficient cross-layer detail
% and low profiling overhead.}
\par\addvspace{4pt}
\noindent\emph{Challenge 2: Combine rich full-stack coverage with low overhead,
a tradeoff existing always-on profiling cannot resolve.}
\par\addvspace{4pt}
\para{Engine-coupled instrumentation.}
Profiling diverse MaaS services cannot depend on application owners changing
source code or startup configurations. PyTorch Profiler, Dynolog, and
StriaTrace require application-side integration or predefined profiling
support~\cite{pytorchprofiler,dynolog,kineto,striatrace}.
As optimization questions change, new collectors or additional functions and
fields may fall outside this support, requiring application changes or service
relaunch.

\par\addvspace{4pt}
\noindent\emph{Challenge 3: Decouple profiling capabilities
and coverage from engine releases and service deployment.}
\par\addvspace{4pt}
\para{Diagnosis alone does not improve performance.}
A finding does not directly determine a safe and effective modification.
Diagnosis systems report suspected bottlenecks~\cite{striatrace,flare}, leaving
implementation and validation to engineers.
Across models, accelerators, and serving conditions, the same inefficiency mechanism can
require different implementations and tests. Experts cannot exhaustively
reproduce and optimize all these combinations manually.

\par\addvspace{4pt}
\noindent\emph{Challenge 4: Translate inefficiency findings into
deployable improvements without extensive expert intervention.}
\par\addvspace{4pt}

\subsection{Design Principles}
\label{sec:motivation-runtime-attachment}

% \para{Key Insight 1: Complex serving conditions require adaptive profiling for full-stack optimization.}
% Predefined instrumentation probes can miss
% relevant state and dependencies; exhaustive always-on tracing can incur
% substantial online overhead. Full-stack optimization needs
% detailed observations whose scope can change with the investigation.

\para{Key Insight 1: Complex serving conditions require adaptive
full-stack observation, not always-on collection.}
Predefined instrumentation probes can miss
relevant state and dependencies; exhaustive always-on tracing can incur
substantial online overhead. Full-stack optimization needs
detailed observations whose scope can change with the investigation.

\para{Key Insight 2: Serving conditions vary, but inefficiency mechanisms and optimization strategies recur.}
LLM inference follows a common structure of scheduling, coordination,
and operator execution. Reference fixes in inference engines and operator
implementations reveal recurring causes of avoidable cost and corresponding
remedies, such as asynchronous execution to avoid unnecessary
stalls, better operator dataflow to reduce memory traffic, and implementations
suited to the target device (\S\ref{sec:analyzer-patterns}).
The mechanism provides a basis for recognizing a current problem; the
reference fix suggests how to address it.
% The current engine,
% workload, and hardware determine whether and how that solution applies.

\vspace{.05in}
\noindent These insights motivate three design principles:
\begin{enumerate}[itemsep=4pt,parsep=0pt,topsep=4pt,partopsep=0pt]
\renewcommand{\labelenumi}{(\arabic{enumi})}
\item \emph{Runtime Attachment for On-Demand Profiling
within Bounded Windows.}
Profiling should follow investigation needs without engine changes
or restarts. Each session should collect the required full-stack detail, then stop;
later sessions target unresolved questions.
\item \emph{Mechanism-Based Analysis and Suggested Solutions.}
Each pattern should associate required evidence and applicability
conditions with reference fixes, so suggestions depend on verified
conditions in the current execution.
\item \emph{Agent-Driven Optimization with Controlled Experiments.}
Agent-driven experiments should preserve triggering conditions and
dependencies at reduced scale, with iterative implementation and testing
followed by expert review of correctness and deployment applicability.
\end{enumerate}

%% file: figs/motivation-ab-figure.tex
% Joint Motivation figure; original wide diagram is retained separately.
\begin{figure}[t]
  \centering
  \subfloat[Cross-component waiting in the KV-service blocking case.\label{fig:eval-kvcache-evidence}]{%
    \begin{minipage}[t]{.485\columnwidth}
    \centering
    \input{figs/eval-kvcache-cross-process-styles}
    \resizebox{\linewidth}{!}{\input{figs/kv-wait-square}}%
    \end{minipage}}%
  \hfill
  \subfloat[Online distributions of GDN input shapes and SOL attainment.\label{fig:motivation-shape-efficiency}]{%
    \begin{minipage}[t]{.485\columnwidth}
    \centering
    \includegraphics[width=\linewidth]{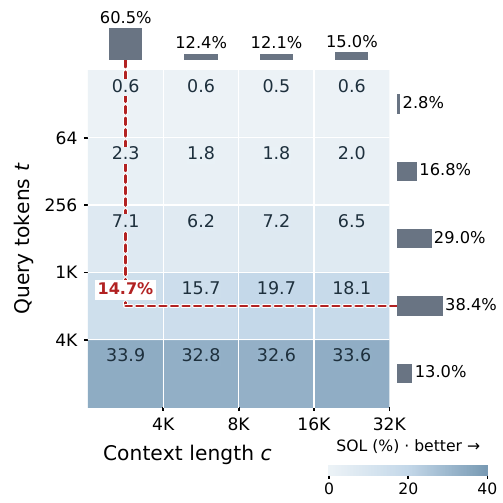}%
    \end{minipage}}
  \caption{Optimization depends on serving conditions.}
  \label{fig:motivation-conditions-ab}
\end{figure}

%% file: figs/eval-kvcache-cross-process-styles.tex
\definecolor{kvhostblue}{RGB}{40,165,194}
\definecolor{kvrpcorange}{RGB}{244,145,41}
\definecolor{kvgpugold}{RGB}{218,169,0}
\definecolor{kvwaitred}{RGB}{214,69,65}
\definecolor{kvguardgreen}{RGB}{49,139,112}
\definecolor{kvink}{RGB}{30,48,61}
\definecolor{kvpanelbg}{RGB}{247,249,250}
\tikzset{
  kvcachefigure/.style={
    x=1cm,y=1cm,
    font=\fontsize{7.1}{8.2}\selectfont,
    text=kvink,
    >={Stealth[length=3.5pt,width=3.2pt]}
  },
  kvflow/.style={->,semithick,draw=kvink!80},
  kvlane/.style={font=\fontsize{6.9}{7.8}\selectfont\itshape,anchor=east},
  kvblock/.style={draw=kvink!70,semithick,rounded corners=0.8pt,
    align=center,inner sep=1.5pt},
  kvwait/.style={kvblock,draw=kvwaitred,fill=kvwaitred!12,
    text=kvwaitred!65!black,dashed},
  kvhost/.style={kvblock,fill=kvhostblue!88,text=white},
  kvrpc/.style={kvblock,fill=kvrpcorange!88,text=black},
  kvgpu/.style={kvblock,fill=kvgpugold!88,text=black},
  kvguard/.style={kvblock,draw=kvguardgreen,fill=kvguardgreen!13,
    text=kvguardgreen!55!black}
}

%% file: figs/kv-wait-square.tex
% Half-column square copy. The original wide figure is preserved unchanged.
% Control-flow schematic: positions and lengths do not encode event duration.
\begin{tikzpicture}[x=1cm,y=1cm,
  font=\sffamily\fontsize{7.0}{7.8}\selectfont,text=kvink,
  >={Stealth[length=3.4pt,width=3.0pt]}]
\definecolor{kvwaitred}{HTML}{B22222}
\definecolor{kvink}{HTML}{202830}
\definecolor{kvhostblue}{HTML}{4E79A7}
\definecolor{kvrpcorange}{HTML}{F28E2B}
\definecolor{kvgpugold}{HTML}{59A14F}
\definecolor{kvengineedge}{HTML}{4E79A7}
\definecolor{kvserviceedge}{HTML}{F28E2B}
\definecolor{kvgpuedge}{HTML}{59A14F}
\tikzset{
  sqbox/.style={draw=kvink!60,line width=.65pt,rounded corners=.6pt,
    align=center,inner sep=1pt},
  sqhost/.style={sqbox,draw=kvengineedge,fill=kvhostblue!20},
  sqrpc/.style={sqbox,draw=kvserviceedge,fill=kvrpcorange!20},
  sqgpu/.style={sqbox,draw=kvgpuedge,fill=kvgpugold!20},
  sqwait/.style={sqbox,draw=kvwaitred,fill=kvwaitred!9,dashed,
    text=kvwaitred!65!black},
  sqflow/.style={->,line width=.7pt,draw=kvink!80},
  sqlabel/.style={font=\sffamily\fontsize{6.3}{7.2}\selectfont\bfseries,anchor=west},
  sqnote/.style={font=\sffamily\fontsize{6.6}{7.4}\selectfont,text=kvink}
}
\path[use as bounding box] (0,0) rectangle (4.15,-4.15);
\draw[draw=kvengineedge,dashed,line width=.65pt,fill=kvhostblue!3,
  rounded corners=.7pt] (.02,-.03) rectangle (4.13,-2.57);
\node[sqnote,anchor=north] at (2.075,-.09) {Supplementary profile};
\node[sqlabel] at (.16,-.72) {KV service};
\node[sqrpc,minimum width=1.69cm,minimum height=.53cm]
  (create) at (2.74,-.72) {allocate KV\\metadata};
\node[sqlabel] at (.16,-1.24) {EngineCore};
\node[sqhost,minimum width=.97cm,minimum height=.40cm]
  (schedule) at (.67,-1.72) {schedule};
\node[sqwait,minimum width=1.39cm,minimum height=.55cm]
  (wait) at (2.14,-1.72) {wait for\\allocation};
\node[sqhost,minimum width=1.01cm,minimum height=.39cm]
  (enqueue) at (3.44,-2.27) {enqueue};
\draw[sqflow] (schedule.east)--(wait.west);
\draw[sqflow] (wait.east) -| (enqueue.north);
\draw[<->,line width=.7pt,draw=kvserviceedge]
  (wait.north) -- (2.14,-1.01) -| (create.south);
\node[sqnote,anchor=west,text=kvink,font=\sffamily\fontsize{6.1}{7}\selectfont]
  at (2.92,-1.18) {sync RPC};

\draw[draw=kvink!70,line width=.65pt,fill=kvpanelbg,
  rounded corners=.7pt] (.02,-2.64) rectangle (4.13,-4.12);
\node[sqnote,anchor=west,font=\sffamily\fontsize{6.0}{7}\selectfont] at (.16,-2.79) {Worker/GPU evidence};
\node[sqlabel] at (.16,-3.20) {Worker};
\node[sqlabel] at (.16,-3.82) {GPU};
\node[sqwait,minimum width=1.55cm,minimum height=.43cm]
  (dequeue) at (1.92,-3.20) {\fontsize{6.3}{7}\selectfont\texttt{dequeue} wait};
\node[sqgpu,minimum width=1.00cm,minimum height=.46cm]
  (execute) at (3.48,-3.20) {execute\\model};
\node[sqbox,fill=white,minimum width=1.55cm,minimum height=.42cm]
  (idle) at (1.92,-3.82) {GPU idle};
\node[sqgpu,minimum width=1.00cm,minimum height=.46cm]
  (compute) at (3.48,-3.82) {model\\execution};
\draw[sqflow] (enqueue.south) -- (dequeue.north east);
\node[sqnote,fill=white,inner sep=.5pt,font=\sffamily\fontsize{6.0}{7}\selectfont] at (3.56,-2.77) {\texttt{workId}};
\draw[sqflow] (dequeue.east)--(execute.west);
\draw[sqflow] (execute.south)--(compute.north);
\end{tikzpicture}

%% file: section/overview.tex
\section{\sys Overview}
\label{sec:overview}
 \sys implements these design principles for LLM
inference on NVIDIA GPUs and in-house accelerator devices.
\figref{fig:webb-overview} shows its architecture.
\textbf{Hprof} attaches probes to running processes to collect selected
Python and accelerator execution records without LLM engine changes or
service restarts. The \textbf{control plane} specifies target processes and
collection windows across the fleet; machine-local Hprof daemons execute
these plans (\S\ref{sec:aprof}).
From these records, \textbf{Analyzer} reconstructs executions and
dependencies to check the conditions of inefficiency patterns
(\S\ref{sec:analyzer}).
Unresolved dependencies lead to an expert-approved plan for supplementary
profiling, while confirmed patterns are combined with supporting evidence
and suggested solutions to form \emph{optimization opportunities} for
\textbf{Optimizer} (\S\ref{sec:optimizer}).
Its agents iteratively develop and test changes through an
autoresearch-style loop~\cite{karpathy-autoresearch}.
Expert review determines which changes proceed to canary validation
and staged rollout outside \sys.

\begin{figure}[t]
  \setlength{\abovecaptionskip}{4pt}
  \centering
  \input{figs/webb-overview-tikz}\par
  \caption{\sys overview.}
  \label{fig:webb-overview}
\end{figure}

%% file: figs/webb-overview-tikz.tex
% Single-column overview, drawn at native manuscript size.
% A--C are schematic deployments, not a measured configuration matrix.
% Previous version: notes/成形模块/系统模块/Overview-单栏缩图-20260909/before-tikz.tex
\begingroup
\definecolor{ovblue}{RGB}{219,232,244}
\definecolor{ovdaemon}{RGB}{173,203,230}
\definecolor{ovgreen}{RGB}{226,238,220}
\definecolor{ovgold}{RGB}{252,242,215}
\definecolor{ovline}{RGB}{77,87,96}
\definecolor{ovattach}{RGB}{72,111,146}
\begin{tikzpicture}[
  x=1pt,y=-1pt,
  font=\sffamily\fontsize{7}{8}\selectfont,
  every node/.style={align=center,inner sep=0pt,outer sep=0pt},
  title/.style={font=\sffamily\bfseries\fontsize{8.2}{9.2}\selectfont},
  heading/.style={font=\sffamily\bfseries\fontsize{6.8}{7.8}\selectfont},
  labeltext/.style={font=\sffamily\fontsize{6.8}{7.8}\selectfont},
  small/.style={font=\sffamily\fontsize{6.7}{7.7}\selectfont},
  probe/.style={font=\sffamily\fontsize{6}{6.8}\selectfont},
  box/.style={draw=ovline,line width=.45pt,fill=white},
  flow/.style={-{Stealth[length=3pt,width=2.5pt]},draw=ovline,line width=.6pt},
  wire/.style={draw=ovline,line width=.6pt},
  attachment/.style={-{Stealth[length=2.2pt,width=1.8pt]},draw=ovattach,line width=.5pt},
  control/.style={flow,dashed,draw=ovattach}
]
\path[use as bounding box] (0,0) rectangle (240,207);

% Collection control and the runtime attachment mechanism.
\draw[box,fill=black!3] (21,1) rectangle (95,29);
\node[title] at (58,10) {Control Plane};
\node[small] at (58,22) {Processes + fields};
\draw[box,fill=ovblue] (110,1) rectangle (230,29);
\node[title] at (170,10) {Hprof};
\node[small] at (170,22) {Runtime-attached profiling};
\draw[control] (95,15)--(110,15);
\draw[control] (170,29)--(170,46);
\node[small,anchor=east,text=ovattach] at (163,37.5)
  {\bfseries On-demand Attachment};

% Fleet-wide scope is explicit; A expands the two process roles.
\draw[box] (21,46) rectangle (230,127);
\node[title,anchor=west] at (26,56) {Fleet-wide Profiling};
\node[probe,anchor=east] at (225,56) {Models\enspace$\cdot$\enspace Engines\enspace$\cdot$\enspace Accelerators};
\draw[box,fill=black!3] (25,65) rectangle (127,121);
\node[labeltext] at (76,71) {Deployment A};
\draw[box] (29,82) rectangle (70,109);
\node[labeltext] at (49.5,90) {EngineCore};
\draw[box,fill=ovblue] (31,98) rectangle (68,107);
\node[probe] at (49.5,102.5) {Python probe};
\draw[box] (76,82) rectangle (125,109);
\node[labeltext] at (100.5,88) {Worker};
\draw[box,fill=ovblue] (78,94) rectangle (123,101);
\node[probe] at (100.5,97.5) {Python probe};
\draw[box,fill=ovblue] (78,102) rectangle (123,109);
\node[probe] at (100.5,105.5) {Accelerator probe};
\draw[flow] (70,89)--(76,89);
\draw[box,fill=ovgold] (76,114) rectangle (125,121);
\node[probe] at (100.5,117.5) {NVIDIA GPUs};
\draw[flow] (100.5,109)--(100.5,114);
% The node-local daemon attaches to the probes in both processes.
\draw[box,fill=ovdaemon] (29,112) rectangle (69,120);
\node[probe,font=\sffamily\bfseries\fontsize{6}{6.8}\selectfont] at (49,116) {Hprof daemon};
\draw[attachment] (49,112)--(49,107);
\draw[attachment] (69,116)--(73,116)--(73,101.5)--(78,101.5);

\foreach \x/\deployment/\hardware in {132/B/XPU-C,181/C/XPU-C}{
  \begin{scope}[shift={(\x,0)}]
    \draw[box,fill=black!3] (0,65) rectangle (45,121);
    \node[labeltext] at (22.5,71) {Deployment \deployment};
    % Repeated replica--accelerator pairs; each replica contains its daemon.
    \foreach \dx/\dy in {4/-4,2/-2,0/0}{
      \begin{scope}[shift={(\dx,\dy)}]
        \draw[box] (4,84) rectangle (37,105);
        \node[labeltext] at (20.5,91) {Replica};
        \draw[box,fill=ovdaemon] (6,97) rectangle (29,103);
        \node[probe] at (17.5,100) {daemon};
      \end{scope}
    }
    % Three visible connections terminate at the matching accelerator cards.
    \foreach \dx/\dy in {4/-4,2/-2,0/0}{
      \begin{scope}[shift={(\dx,\dy)}]
        \draw[flow] (36,105)--(36,114);
        \draw[box,fill=ovgold] (4,114) rectangle (37,121);
      \end{scope}
    }
    \node[probe] at (20.5,117.5) {\hardware};
  \end{scope}
}

% Show one representative records path; omit duplicate links from B and C.
\draw[flow] (49,120)--(49,125)--(67,125)--(67,145);
\node[small,anchor=west] at (75,136) {PROFILING RECORDS};
\draw[box,fill=ovgreen] (21,145) rectangle (203,165);
\node[title,anchor=west] at (29,155) {Analyzer};
% A schematic execution graph is matched against patterns and solutions.
% The arrow denotes a lookup/match, not insertion into the database.
\draw[draw=ovline,line width=.45pt]
  (71,151)--(80,148.5)--(88.5,152)
  (71,151)--(79,155.5)--(88.5,152)
  (79,155.5)--(71.5,161)
  (79,155.5)--(87,161)--(89.5,157)--(88.5,152);
\foreach \x/\y in {71/151,80/148.5,88.5/152,79/155.5,71.5/161,87/161,89.5/157}{
  \draw[draw=ovattach,fill=white,line width=.5pt]
    (\x,\y) circle [radius=1.6pt];
}
\draw[flow] (96,155)--(125,155);
% A magnifying glass over the matching arrow echoes the linked observations.
\draw[draw=ovline,fill=white,line width=.7pt]
  (108,153) circle [radius=4.2pt];
\draw[draw=ovline,line width=.9pt,line cap=round]
  (111,156)--(115,160);
\draw[draw=ovattach,line width=.45pt]
  (105.8,153)--(108,151.2)--(110,154);
\foreach \x/\y in {105.8/153,108/151.2,110/154}{
  \fill[ovattach] (\x,\y) circle [radius=.65pt];
}
% Two separate knowledge sources share the compact database symbol.
\path[fill=ovgold] (130,149) rectangle (141,160);
\draw[draw=ovline,line width=.5pt]
  (130,149)--(130,160)
  .. controls (130,163) and (141,163) .. (141,160)--(141,149);
\draw[draw=ovline,fill=ovgold,line width=.5pt]
  (135.5,149) ellipse [x radius=5.5pt,y radius=2pt];
\draw[draw=ovline,line width=.4pt]
  (130,154).. controls (130,157) and (141,157) .. (141,154);
\node[small,anchor=west] at (145,155) {Inefficiency Patterns\\Reference Fixes};
\draw[flow] (67,165)--(67,187);
\node[small,anchor=west] at (75,176) {OPTIMIZATION OPPORTUNITIES};
\draw[box,fill=ovblue] (21,187) rectangle (203,207);
\node[title,anchor=west] at (29,197) {Optimizer};
% Iterative code changes and tests precede expert review.
\draw[draw=ovline,fill=white,line width=.55pt]
  (78,192) rectangle (93,202);
\draw[draw=ovline,line width=.55pt,line cap=round,line join=round]
  (83,195)--(80.5,197)--(83,199)
  (88,195)--(90.5,197)--(88,199)
  (86.5,194.5)--(84.5,199.5);
\draw[draw=ovattach,line width=.5pt,-{Stealth[length=2.4pt,width=2pt]}]
  (76,194).. controls (76,187.5) and (92,187.5) .. (95,192);
\draw[draw=ovattach,line width=.5pt,-{Stealth[length=2.4pt,width=2pt]}]
  (95,200).. controls (95,206.5) and (79,206.5) .. (76,202);
\node[small,anchor=west] at (98,197) {Agent-driven\\Optimization loop};
\draw[flow] (150,197)--(157,197);
\draw[draw=ovline,fill=white,line width=.55pt] (166,192) circle [radius=2.6pt];
\draw[draw=ovline,fill=white,line width=.55pt]
  (160,203).. controls (160,195) and (172,195) .. (172,203)--cycle;
\node[small,anchor=west] at (178,197) {Expert\\Review};

% Supplementary investigation returns to collection control.
\draw[control] (21,155)--(12,155)--(12,15)--(21,15);
\node[small,rotate=90,text=ovattach] at (4.5,87)
  {Supplementary Investigation};

% The external release path runs along the right-hand side.
\draw[box,fill=black!3,dashed] (219,136) rectangle (239,207);
\node[heading,rotate=90] at (225,171.5) {Canary and Rollout};
\node[probe,rotate=90] at (234,171.5) {\emph{outside Herschel}};
\draw[flow] (203,197)--(219,197);
\draw[flow] (229,136)--(229,127);
\end{tikzpicture}
\endgroup

%% file: section/aprof.tex
\section{Hprof: Fleet-Wide On-Demand Profiling}
\label{sec:aprof}
\label{sec:wprof}

\begin{ourrevision}
Hprof collects Python calls, runtime object fields, host API calls, and
accelerator activities. It attaches probes at runtime
and manages their  lifecycle through an in-process
controller.

\subsection{Runtime-Attached Profiling}
\label{sec:wprof-runtime}

The \textbf{control plane} converts collection requests into process-specific
plans (\S\ref{sec:wprof-services}). Each plan specifies probes, selected
calls, supported object fields, and a recording window. A Hprof daemon
process runs on each machine and invokes the CLI to execute these plans.
In \figref{fig:wprof-architecture}, the CLI establishes an in-process
controller and sends it commands; the controller configures
and operates probes through \emph{Probe control}.

\begingroup
\para{Attachment.}
\clubpenalty=10000
Hprof uses thread redirection only for initial entry into a worker process;
subsequent collection commands reach the controller through \emph{Socket
control}. As shown in \figref{fig:wprof-architecture}, the CLI pauses a
selected thread with \texttt{ptrace} and inspects its stack,
excluding known allocator, loader, and threading contexts that could hold
locks needed by injected calls. It temporarily redirects the thread to call
\texttt{pthread\_create}, creating a loader thread that loads the in-process
controller library with \texttt{dlopen}. The CLI restores the application
thread and releases \texttt{ptrace} before waiting for loading to finish.
\par\endgroup

\para{Session control.}
Through \emph{Probe control}, the controller applies each session plan's
probe configuration and starts recording, which proceeds independently of
command processing. A stop request or session-lease expiry stops recording and starts draining
queued records. \S\ref{sec:wprof-collection} details
probe-specific collection and stopping.

\para{Resource reclamation and reuse.}
Probe resources can outlive a recording session: outstanding calls or
callbacks may still reach their code. Each probe therefore reports
whether its resources are safe to reclaim or reuse. The controller
unloads a probe only when reclamation is safe and permits reuse only when the
retained state is reusable. 
% Probes enforce window boundaries so work from an earlier session cannot
% publish records into a later one.

\begin{figure}[t]
  \setlength{\abovecaptionskip}{4pt}
  \centering
  \includegraphics[width=\columnwidth]{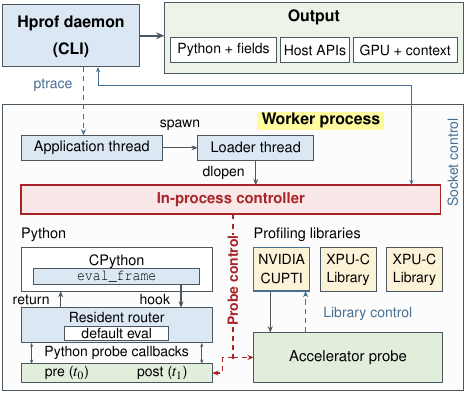}\par
  \color{ourrevisioncolor}
  \caption{Hprof collection and delivery.}
  \label{fig:wprof-architecture}
\end{figure}

\subsection{Cross-Layer Profiling Collection}
\label{sec:wprof-collection}

In \figref{fig:wprof-architecture}, the Python probe records selected engine
calls and supported object fields; the accelerator probe records host API
calls and device activity.

\para{Python calls and object fields.}
The Python probe uses a resident router to intercept CPython frame
evaluation. Callbacks around the default evaluator record a selected
call's function, native thread, and entry/return times; entry callbacks
also capture supported object fields. 
The interval includes native execution and waiting, not just interpreter
CPU time. Engine-specific readers extract tensor dimensions and token counts
to distinguish batches at the same model-forward entry, retaining links
to the originating calls. New object layouts require updating the corresponding Hprof reader, independent of the engine release.
Early filtering and shared symbol definitions reduce per-call work;
bounded queues defer output. Field readers preserve
exception state and suppress profiling during field extraction. Recording
resumes for normal application execution, including nested calls.

\para{Host API calls and device activity.}
Accelerator probes use vendor profiling libraries. On NVIDIA, CUPTI
(CUDA Profiling Tools Interface)~\cite{nvidia-cupti} reports CUDA Runtime
API intervals and activity for kernels, memory operations, and
synchronization. Host records describe submissions and waits; device
records describe work that can continue after the enclosing call returns.
\texttt{correlation id} associates submissions with device activity;
device and stream identifiers locate the work.
% Host coverage is limited to accelerator APIs, rather than arbitrary C/C++ functions.

\para{Heterogeneous Devices Beyond NVIDIA.}
Hprof connects vendor tracing libraries to its common probe interface
through adapters. The AMD backend, for example, uses
ROCTracer~\cite{amd-roctracer}, while the in-house XPU-C backend uses
CUPTI-like APIs. Supporting a new device may require targeted updates to the relevant adapter, reader,
or reconstruction rule.

% \para{Session transitions.}
% The resident router keeps outstanding call returns outside unloadable
% probe code. Stopping closes its callback entry and restores the interpreter
% hook; calls already in the default evaluator continue through the router,
% which invokes return callbacks only if their original session remains active.
% On NVIDIA, completed activity buffers enter a bounded writer queue for
% deferred processing. To enable or disable activity collection, the backend gates new
% callers at CUDA API entry, waits for tracked API calls to return, changes
% state, and releases callers. This \emph{Library control} gate covers
% transitions, not the recording window. Required resources remain resident
% if CUPTI cannot yet be finalized. Filtering each activity record by its
% start time separates successive windows despite delayed buffer delivery.

\para{Session transitions.}
The resident router keeps in-flight returns outside unloadable probe code.
Stopping closes its callback entry and restores the interpreter hook.
Admitted default-evaluator calls return through the router; callbacks run
only for active originating sessions. Completed NVIDIA activity buffers
enter a bounded writer queue. Collection transitions block CUDA API entry,
drain tracked calls, switch state, and release callers. The
\emph{Library control} gate excludes recording; resources stay resident
pending CUPTI finalization. Start-time filtering separates windows despite
delayed delivery.

% \para{A profile bundle for Analyzer.}
% \label{sec:wprof-evidence}
% The daemon delivers these outputs---Python calls with
% fields, host API calls, and GPU activity with device context---to
% Analyzer (\S\ref{sec:analyzer-evidence}). The bundle retains original
% PID/TID, available correlation identifiers,
% and shared deployment, configuration, collection-window, and available
% loss or truncation information.
% If Analyzer needs additional observations, an expert-approved plan
% revision requests supplementary profiling through the same control path.

\subsection{Fleet-Wide Profiling Coordination}
\label{sec:wprof-services}

\para{Fleet-wide coordination.}
Collection requests specify the serving conditions to investigate
(\S\ref{sec:motivation}). The control plane applies deployment-specific
sampling policies to select instances and recording times across
models and accelerator types. Hprof daemons report process identities,
roles, and device associations before the control plane dispatches
plans to selected processes. Canary collections and concurrency
limits constrain collection across instances.

\para{Bounded collection window.}
For initial investigations of sustained inefficiencies, Hprof typically
collects for 30 s. In production traces from our typical workloads,
each window captured at least ten complete inference steps, with a
median of over 700. This supports observing repeated execution
behavior within a short collection session.

% Delivery discussion is being moved to Section 5.2.
% \para{Delivering evidence to Analyzer.}
% \label{sec:wprof-evidence}
% Hprof daemons upload a profile bundle for Analyzer containing Python calls
% and selected fields, host API intervals, and device activity
% (Figure~\ref{fig:wprof-architecture}). The bundle also carries deployment,
% process, configuration, recording-window, and collection-quality metadata.
% Original thread and timestamp references
% are retained even when display identifiers are remapped; event coverage,
% captured fields, and available identifiers depend on backend support and
% the collection plan. Analyzer uses these
% records to recover operator executions, their input properties, and
% engine dependencies (\S\ref{sec:analyzer-evidence}). Collection scope
% and loss information qualify these observations. If a required dependency
% or runtime state remains unobserved, an expert-approved plan revision
% requests additional processes, functions, or supported fields through
% the same collection path (\S\ref{sec:analyzer-patterns}).

\end{ourrevision}

%% file: section/analyzer.tex
% Declare the wide observation figure early so it appears on page 6.
\input{figs/analyzer-observations-row}

% Queue the pattern table with Figure 5 on page 6.
\begin{table*}[!t]
\centering
\caption{Inefficiency patterns and historical PR/CR counts.}
\label{tab:analyzer-categories}
\input{figs/analyzer-categories-table}
\end{table*}

\section{Analyzer: From Runtime Evidence to Optimization Opportunities}
\label{sec:analyzer}
Analyzer recovers executions from Hprof profiles, structures performance observations, and associates reference fixes with confirmed mechanisms. Together, these results form optimization opportunities that guide Optimizer’s experiments.

% \begin{figure}[t]
%   \centering
%   \input{figs/analyzer-overview-no-records-tikz}
%   \caption{Analyzer workflow.}
%   \label{fig:analyzer-evidence}
% \end{figure}

\subsection{Deriving Performance Observations}
\label{sec:analyzer-evidence}
\label{sec:analyzer-records}
\label{sec:analyzer-observations}

\figref{fig:analyzer-cross-layer} summarizes Analyzer's operator measurements
and observations of execution progress and state.

\para{Operator execution and efficiency.}
Measuring operator efficiency requires real inputs and a complete execution.
Implementation-specific rules recover each invocation's \emph{kernel
composition} from call-to-kernel links (\S\ref{sec:wprof-collection}). Because
\emph{CUDA Graph replay} hides operator calls, graph identities and launch
correlations separate replays; model-specific rules recover operators along the
layer sequence, and submission time binds current inputs. Missing composition
or inputs leave the measurement unresolved.

In \figref{fig:analyzer-operator-recovery}, $T_{\mathrm{op}}$ spans the first
kernel start to the last completion. Analyzer recovers shapes and request
lengths from Python object fields, model configuration, and sharding. With data
types and semantics, these yield theoretical operation counts and logical data
volumes; Graph padding is tracked separately. Normalizing both by
$T_{\mathrm{op}}$ and the corresponding device peaks gives model FLOPs
utilization (MFU) and model bandwidth utilization
(MBU)~\cite{llm-inference-performance2023}. We report speed-of-light (SOL)
attainment against the compute/bandwidth bound as
\mbox{$\mathrm{SOL}_{\mathrm{op}}=\max(\mathrm{MFU},\mathrm{MBU})$}~\cite{solar2026,sol-execbench,roofline2008}.

Efficient execution can still produce unneeded results. The \emph{dataflow}
in \figref{fig:analyzer-cross-layer-fusion}, verified against executed code
and supplemented by available object identities, exposes materialization
and external uses~\cite{xla-gpu-architecture}, constraining which work
later optimization may remove.

\para{Execution progress and state.}
Local measurements alone do not explain step delays. From phase timings and
GPU activity, Analyzer separates communication hidden by computation from
exposed time, as shown in \figref{fig:analyzer-cross-layer-overlap}~\cite{hta}.
Submission, execution-order, and synchronization dependencies identify the
\emph{critical path} in \figref{fig:analyzer-cross-layer-critical-path}; missing
dependencies leave delay contributions unresolved~\cite{hta-critical-path}.
Shared call nesting can trace the cross-process
wait in \figref{fig:analyzer-cross-layer-process} to synchronous RPCs delaying
dispatch~\cite{striatrace}. For untraced services, only caller-side activity is visible.
Retaining state may avoid repeated work at the cost of memory. Analyzer checks
state validity and actual \emph{state reuse}; allocation, use, and release establish
\emph{GPU memory lifetimes}, showing memory held between uses.

\subsection{Characterizing Inefficiency Patterns}
\label{sec:analyzer-patterns}
\label{sec:analyzer-pattern-sources}
\para{Recurring inefficiency patterns.}
As noted in Key Insight~2 (\S\ref{sec:motivation-runtime-attachment}),
inefficiency mechanisms and optimization strategies recur despite changing
serving conditions. To capture this reusable knowledge, we examined over
20,000 performance-related PR/CR records from open-source
projects~\cite{vllm-repository,sglang,tensorrtllm,flashinfer}. Drawing on these records, expert knowledge, and
production investigations, we grouped recurring causes by mechanism
and associated them with reference fixes. \tabref{tab:analyzer-categories}
summarizes 35 patterns: 25 engine patterns across eight
categories and 10 operator patterns across three categories.

\para{Inefficiencies in inference engines.}
\label{sec:analyzer-engine-patterns}
% The analysis depends on the conditions that make each behavior wasteful.
For waits, Analyzer compares readiness, dispatch, and first use: unfinished
prerequisites direct investigation upstream, whereas ready work dispatched
late implicates submission. Ready, independent stages executed serially
suggest overlap. For repeated work, state identity and validity determine
whether reconstruction could have reused a compatible result. An output
without a consumer may be removable, but its computation can still perform
a required state update. Speculative decoding is assessed by combined
drafting and verification cost per generated token, not acceptance alone.
Optimizer tests whether a change saves time after accounting for
contention, retained memory, and Graph-padding costs.

\para{Inefficiency patterns in operator implementations.}
\label{sec:analyzer-operator-patterns}
Operator efficiency can vary sharply with input dimensions, even for an
implementation tuned offline. We therefore compare the recovered operator
$\mathrm{SOL}_{\mathrm{op}}$ with state-of-the-art (SOTA) reference targets
matched for semantics, precision, and shapes. Targets come from workload-specific
benchmarks~\cite{atrex-bench,sol-execbench}, operator
leaderboards~\cite{sol-execbench-leaderboard,flashinfer-bench-leaderboard}, or optimized
libraries~\cite{nvidiaCublas,nvidiaCutlassEfficientGemm}.
For new accelerators, a well-optimized reference GPU supplies an initial
$\mathrm{SOL}_{\mathrm{op}}$ target. A gap directs attention to implementation selection and
unnecessary data movement. Fusion can avoid materialization; splitting can
expose parallelism. Both require comparing complete operators, including
their added or removed stages.
To compute \emph{Hotness}, we weight each shape's estimated recoverable
fraction at the reference target by its share of the workload's critical-path
time and sum these contributions within each workload. We then weight the
workload totals by their shares of fleet GPU-hours over the same period. It ranks opportunities, not validated gains.

\para{Matching observations and supplementary profiling.}
\label{sec:analyzer-supplementary}
% A pattern match binds its calls, objects, and dependencies to the current
% execution and checks every required condition. Optional evidence adds
% context; counterevidence rejects a match. Missing evidence leaves the match unresolved and triggers
% \emph{supplementary profiling}. For example,
% establishing that command receipt gates a Worker's execution explains its
% local wait but not the upstream delay. Additional EngineCore calls and
% shared work ids distinguish preparation blocked upstream from delayed
% dispatch, as in \S\ref{sec:eval-kvcache}. Analyzer requests the missing
% processes, functions, and supported fields through an expert-approved plan;
% Hprof collects them in a bounded window to resolve the relationship.
A pattern specifies the conditions under which an observed behavior
incurs avoidable cost. Matching relates the pattern's calls, objects, and
dependencies to current executions. 
Object identities and code semantics establish whether
the work is required and when its results or state updates are needed.
A conflicting condition rejects the match. Missing required evidence
leaves the mechanism unresolved; supplementary profiling targets
relationships that additional runtime records can establish.
% supplementary
For example, we revisit the KV case in
\figref{fig:eval-kvcache-evidence}, where Workers wait for execution
commands while GPUs remain idle. The initial evidence locates the wait
but does not explain EngineCore's delayed dispatch. Analyzer therefore
requests EngineCore calls and shared work IDs through an expert-approved
plan. Hprof collects these records in another bounded window, linking the Worker wait to synchronous object creation before
dispatch (\S\ref{sec:eval-kvcache}).

% supplementary

\para{Reference fixes.}
\label{sec:analyzer-identification}
A confirmed mechanism provides a basis for selecting a reference fix;
the current execution determines how it can be applied. Analyzer
retains the fix's source, code changes, and applicability, and passes
suggested solutions with the supporting evidence to Optimizer.
An asynchronous KV-creation fix, for example, must ensure that object
creation completes before KV saving. Optimizer implements the change
and tests correctness and serving performance under current serving conditions.

%% file: figs/analyzer-observations-row.tex
% One double-column row, ordered by first citation in Analyzer.
% Original operator figure and four-panel figure remain in their source files.
\begin{figure*}[t]
  \setlength{\abovecaptionskip}{4pt}
\centering
\begingroup
\input{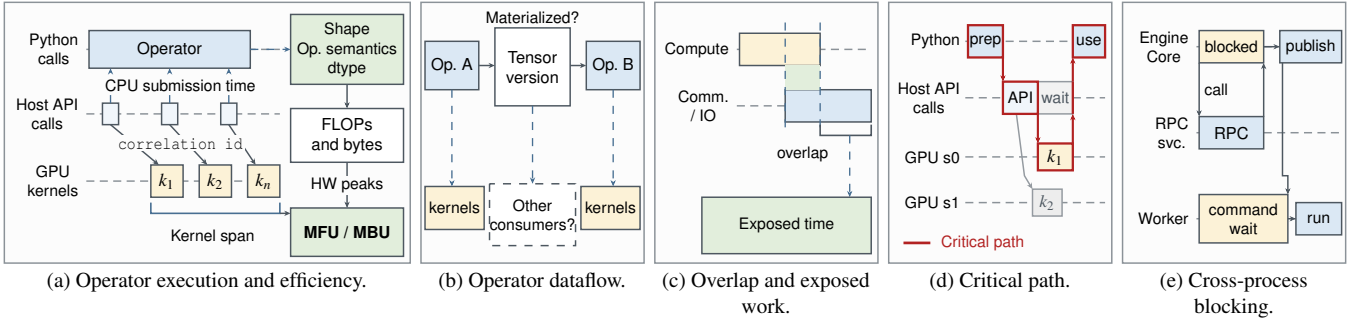}
\tikzset{obsrow/.style={cldiagram,
  font=\fontfamily{phv}\fontsize{6.8}{7.4}\selectfont,
  clsmall/.style={font=\fontfamily{phv}\fontsize{6.5}{7.1}\selectfont},
  clhead/.style={font=\fontfamily{phv}\bfseries\fontsize{6.8}{7.4}\selectfont},
  clflow/.style={-{Stealth[length=2.4pt,width=2pt]},draw=clline,line width=.55pt},
  classoc/.style={-{Stealth[length=2.4pt,width=2pt]},draw=clattach,dashed,line width=.5pt}
}}
\captionsetup[subfloat]{font=footnotesize,labelfont=normalfont,
  justification=centering,singlelinecheck=false,captionskip=2pt,
  farskip=0pt,nearskip=0pt}
\subfloat[Operator execution and efficiency.\label{fig:analyzer-operator-recovery}]{%
\resizebox{.304\textwidth}{!}{%
\begin{tikzpicture}[obsrow]
\path[use as bounding box] (0,0) rectangle (152,98);
\draw[clpanel] (.5,.5) rectangle (151.5,97.5);
\foreach \yy in {18,42,67}{\draw[cllane] (31,\yy)--(103,\yy);}
\node[clsmall,anchor=east] at (28,18) {Python\\calls};
\node[clsmall,anchor=east] at (28,42) {Host API\\calls};
\node[clsmall,anchor=east] at (28,67) {GPU\\kernels};
\draw[clbox,fill=clblue] (32,11) rectangle (92,25);
\node at (62,18) {Operator};
\foreach \xx in {40,62,84}{
  \draw[clbox,fill=clblue!40] (\xx-3,38) rectangle (\xx+3,46);
  \draw[classoc] (\xx,38)--(\xx,25);
}
\node[clsmall,fill=black!1,inner sep=.7pt] at (66,31.5) {CPU submission time};
\draw[clbox,fill=clgold] (55,61) rectangle (67,73);
\draw[clbox,fill=clgold] (73,61) rectangle (85,73);
\draw[clbox,fill=clgold] (91,61) rectangle (103,73);
\node at (61,67) {$k_1$};
\node at (79,67) {$k_2$};
\node at (97,67) {$k_n$};
\draw[clflow] (40,46)--(58,61);
\draw[clflow] (62,46)--(76,61);
\draw[clflow] (84,46)--(94,61);
\node[clsmall,fill=black!1,inner sep=.7pt] at (66,53) {\texttt{correlation id}};
% Full member span, without introducing another duration/recovery block.
\draw[clwire,draw=clattach] (55,75)--(55,79)--(103,79)--(103,75);
\node[clsmall] at (78,88) {Kernel span};
\draw[clbox,fill=clgreen] (107,5) rectangle (149,31);
\node[clsmall] at (128,18) {Shape\\Op. semantics\\dtype};
\draw[classoc] (92,18)--(107,18);
\draw[clbox] (107,40) rectangle (149,60);
\node at (128,50) {FLOPs\\and bytes};
\draw[clflow] (128,31)--(128,40);
\draw[clbox,fill=clgreen] (107,77) rectangle (149,95);
\node[clhead] at (128,86) {MFU / MBU};
\draw[clflow] (128,60)--(128,77);
\node[clsmall,fill=black!1,inner sep=.7pt] at (128,68.5) {HW peaks};
\draw[clflow] (103,79)--(107,79);
\end{tikzpicture}}}%
\hfill
\subfloat[Operator dataflow.\label{fig:analyzer-cross-layer-fusion}]{%
\resizebox{.168\textwidth}{!}{%
\begin{tikzpicture}[obsrow]
\path[use as bounding box] (0,0) rectangle (84,98);
\draw[clpanel] (.5,.5) rectangle (83.5,97.5);
\draw[clbox,fill=clblue] (2,15) rectangle (22,33);
\node at (12,24) {Op. A};
\draw[clbox] (28,10) rectangle (56,39);
\node at (42,24.5) {Tensor\\version};
\draw[clbox,fill=clblue] (62,15) rectangle (82,33);
\node at (72,24) {Op. B};
\draw[clflow] (22,24)--(28,24);
\draw[clflow] (56,24)--(62,24);
\node[clsmall] at (42,6) {Materialized?};
\draw[clbox,fill=clgold] (2,69) rectangle (24,85);
\node[clsmall] at (13,77) {kernels};
\draw[clbox,fill=clgold] (60,69) rectangle (82,85);
\node[clsmall] at (71,77) {kernels};
\draw[classoc] (12,33)--(12,69);
\draw[classoc] (72,33)--(72,69);
\draw[clbox,dashed] (26,68) rectangle (58,91);
\node[clsmall] at (42,79.5) {Other\\consumers?};
% Materialization is a tensor property, not a dataflow stage.
\draw[classoc] (42,39)--(42,68);
\end{tikzpicture}}}%
\hfill
\subfloat[Overlap and exposed work.\label{fig:analyzer-cross-layer-overlap}]{%
\resizebox{.168\textwidth}{!}{%
\begin{tikzpicture}[obsrow]
\path[use as bounding box] (0,0) rectangle (84,98);
\draw[clpanel] (.5,.5) rectangle (83.5,97.5);
\node[clsmall,anchor=east] at (28,18) {Compute};
\node[clsmall,anchor=east] at (28,39) {Comm.\\/ IO};
\draw[cllane] (31,18)--(81,18);
\draw[cllane] (31,39)--(81,39);
\draw[clbox,fill=clgold] (32,12) rectangle (62,24);
\draw[clbox,fill=clblue] (49,33) rectangle (81,45);
\path[fill=clgreen] (49,24) rectangle (62,33);
\draw[draw=clattach,dashed,line width=.5pt] (49,8)--(49,50);
\draw[draw=clattach,dashed,line width=.5pt] (62,8)--(62,50);
\node[clsmall] at (55.5,57) {overlap};
\draw[clwire] (62,46)--(62,50)--(81,50)--(81,46);
\draw[classoc] (73,50)--(73,73);
% Exposed time is interval coverage; critical-path impact is shown in (d).
\draw[clbox,fill=clgreen] (18,73) rectangle (81,94);
\node[clsmall] at (49.5,83.5) {Exposed time};
\end{tikzpicture}}}%
\hfill
\subfloat[Critical path.\label{fig:analyzer-cross-layer-critical-path}]{%
\resizebox{.168\textwidth}{!}{%
\begin{tikzpicture}[obsrow]
\definecolor{clcritical}{RGB}{178,34,34}
\tikzset{cpbox/.style={clbox,draw=clcritical,line width=.85pt},
  cpflow/.style={clflow,draw=clcritical,line width=.8pt}}
\path[use as bounding box] (0,0) rectangle (84,98);
\draw[clpanel] (.5,.5) rectangle (83.5,97.5);
\foreach \yy in {15,36,58,75}{\draw[cllane] (29,\yy)--(81,\yy);}
\node[clsmall,anchor=east] at (27,15) {Python};
\node[clsmall,anchor=east] at (27,36) {Host API\\calls};
\node[clsmall,anchor=east] at (27,58) {GPU s0};
\node[clsmall,anchor=east] at (27,75) {GPU s1};
% A synchronization wait is context, not another copy of device cost.
\draw[clbox,fill=black!5,draw=clline!60] (56,30) rectangle (69,42);
\node[clsmall,text=clline] at (62.5,36) {wait};
\draw[clflow,draw=clline!60] (48,42)--(51,65)--(54,70);
\draw[clbox,fill=black!7,draw=clline!60] (54,70) rectangle (65,80);
\node[clsmall,text=clline] at (59.5,75) {$k_2$};
\draw[cpbox,fill=clblue] (30,9) rectangle (43,21);
\node[clsmall] at (36.5,15) {prep};
\draw[cpbox,fill=clblue!40] (43,30) rectangle (56,42);
\node[clsmall] at (49.5,36) {API};
\draw[cpflow] (43,21)--(43,30);
\draw[cpbox,fill=clgold] (56,53) rectangle (69,63);
\node at (62.5,58) {$k_1$};
\draw[cpflow] (56,42)--(56,53);
\draw[cpflow] (69,53)--(69,42);
\draw[draw=clcritical,line width=.85pt] (69,42)--(69,30);
\draw[cpflow] (69,30)--(69,21);
\draw[cpbox,fill=clblue] (69,9) rectangle (81,21);
\node[clsmall] at (75,15) {use};
\draw[draw=clcritical,line width=1pt] (6,90)--(16,90);
\node[clsmall,text=clcritical,anchor=west] at (20,90) {Critical path};
\end{tikzpicture}}}%
\hfill
\subfloat[Cross-process blocking.\label{fig:analyzer-cross-layer-process}]{%
\resizebox{.168\textwidth}{!}{%
\begin{tikzpicture}[obsrow]
\path[use as bounding box] (0,0) rectangle (84,98);
\draw[clpanel] (.5,.5) rectangle (83.5,97.5);
\foreach \yy in {17,49,81}{\draw[cllane] (28,\yy)--(82,\yy);}
\node[clsmall,anchor=east] at (25,17) {Engine\\Core};
\node[clsmall,anchor=east] at (25,49) {RPC\\svc.};
\node[clsmall,anchor=east] at (25,81) {Worker};
\draw[clbox,fill=clgold] (29,11) rectangle (53,23);
\node[clsmall] at (41,17) {blocked};
\draw[clbox,fill=clblue] (29,43) rectangle (53,55);
\node at (41,49) {RPC};
\draw[clflow] (29,23)--(29,43);
\node[clsmall,anchor=west] at (31,33) {call};
\draw[clflow] (53,43)--(53,23);
\draw[clbox,fill=clblue] (59,11) rectangle (82,23);
\node[clsmall] at (70.5,17) {publish};
\draw[clflow] (53,17)--(59,17);
\draw[clbox,fill=clgold] (29,72) rectangle (62,90);
\node[clsmall] at (45.5,81) {command\\wait};
\draw[clbox,fill=clblue] (65,75) rectangle (82,87);
\node[clsmall] at (73.5,81) {run};
% Command reception ends the wait; it does not initiate it.
\draw[clflow] (60,23)--(60,65)--(62,65)--(62,72);
\draw[clflow] (62,81)--(65,81);
\end{tikzpicture}}}%
\endgroup
\caption{Key performance observations.}
\label{fig:analyzer-cross-layer}
\end{figure*}

%% file: figs/analyzer-categories-table.tex
% Three columns; shared linear bar scale, 0--2,000 PRs/CRs.

\begingroup
\fontsize{9}{10.4}\selectfont
\setlength{\tabcolsep}{3pt}
\renewcommand{\arraystretch}{1}
\definecolor{catalogengine}{RGB}{78,111,141}
\definecolor{catalogoperator}{RGB}{102,131,81}
\newcommand{\cataloggroup}[3]{%
  \begingroup\setlength{\fboxsep}{0pt}%
  \colorbox{#1!10}{\makebox[\linewidth][l]{%
    \hspace{2pt}\strut\textbf{#2}\quad #3}}\endgroup}
\newcommand{\catalogbar}[2]{%
  \begin{tikzpicture}[baseline=0pt,x=1pt,y=1pt]
    \path[use as bounding box] (0,0) rectangle (28,5);
    \fill[black!6] (0,0) rectangle (28,5);
    \fill[#2] (0,0) rectangle ({#1/2000*28},5);
  \end{tikzpicture}\hspace{3pt}\makebox[25pt][r]{%
    \pgfmathprintnumber[1000 sep={,},precision=0]{#1}}}
\noindent\begin{tabular}{@{}>{\raggedright\arraybackslash}m{124pt}
  >{\raggedright\arraybackslash}m{311pt}
  >{\raggedleft\arraybackslash}m{56pt}@{}}
\toprule
\textbf{Category} & \textbf{Inefficiency patterns} & \makebox[\linewidth][r]{\shortstack[r]{\textbf{\# historical PRs/CRs}\\from open-source projects~\cite{vllm-repository,sglang,tensorrtllm,flashinfer}}} \\
\midrule
\multicolumn{3}{@{}l@{}}{\cataloggroup{catalogengine}{Inference engines}{8 categories, 25 patterns}} \\
\textbf{Host supply \& scheduling} & Late submission; repeated input preparation; eligible engine mode rejected & \catalogbar{1090}{catalogengine} \\
\textbf{Synchronization} & Early/repeated readback; spin waits; DP synchronization & \catalogbar{276}{catalogengine} \\
\textbf{Graph capture/replay} & Uncaptured eligible work; capture/compile breaks; excess padding & \catalogbar{1954}{catalogengine} \\
\textbf{State reuse} & Missed KV-prefix reuse; repeated construction of metadata, plans, or Graphs & \catalogbar{886}{catalogengine} \\
\textbf{KV \& runtime memory} & Failed/repeated allocation; capacity mismatch & \catalogbar{1146}{catalogengine} \\
\textbf{Data movement \& remote KV} & KV-transfer waits; blocking host staging; synchronous KV metadata allocation & \catalogbar{1826}{catalogengine} \\
\textbf{Comm. \& placement} & Uncovered collective time; collectives blocking scheduling; unsuitable rank placement & \catalogbar{1798}{catalogengine} \\
\textbf{Speculation \& unused work} & Unsuitable draft count; zero token acceptance; outputs with no consumers & \catalogbar{375}{catalogengine} \\
\midrule
\multicolumn{3}{@{}l@{}}{\cataloggroup{catalogoperator}{Operator implementations}{3 categories, 10 patterns}} \\
\textbf{Shape-dependent efficiency} & Low $\mathrm{SOL}_{\mathrm{op}}$ versus reference efficiency; limited launch parallelism & \catalogbar{1389}{catalogoperator} \\
\textbf{Implementation selection} & Unused eligible specialized paths; library limits; poor algorithm choices & \catalogbar{195}{catalogoperator} \\
\textbf{Operator dataflow} & Costly reduction/combine kernels; missed fusion within/across operators & \catalogbar{1461}{catalogoperator} \\
\bottomrule
\end{tabular}
% \par\smallskip
% {\raggedright\noindent Counts cover each full category of historical PRs/CRs, not production occurrences. Categories may overlap; bars share a common scale.\par}
\endgroup

%% file: section/optimizer.tex
\section{Optimizer: Agent-Driven Optimization}
\label{sec:optimizer}
\label{sec:optimizer-agent}
\label{sec:optimizer-expert}

Optimizer turns evidence-supported opportunities into controlled
experiments. Its agents use execution evidence and suggested fixes to
implement and test candidate changes. Expert reviews select changes for
canary validation and rollout.

\para{Controlled-Scale Experiments on Idle GPUs.}
Each experiment defines correctness checks, a performance objective, and iteration budgets. Reduced deployments must retain workload
conditions and dependencies that expose the implicated work or dependency
in the unchanged baseline without a masking bottleneck; hardware-dependent
hypotheses require the target device. Experiments typically use 8--32
accelerators on idle or revocable capacity~\cite{asi,bamboo2023}, including
prefill/decode resources for disaggregated serving. Completed code and
measurements persist across preemption; interrupted experiments are rerun.
Results establish effects at the tested scale; expert review and canary
validation precede wider rollout.

\para{Agent-Driven Auto-Optimization Loop.}
The agent uses available reference fixes and current evidence to propose a
change, then edits, builds, and tests it~\cite{karpathy-autoresearch}.
Build failures, incorrect results, and performance measurements guide further
revisions or lead the agent to abandon an approach. To measure gains, it
compares the candidate with the baseline under identical workloads and
settings, with high-overhead profiling disabled. Unresolved performance
questions can prompt detailed offline profiling~\cite{ncu-profiling-guide}.
The loop stops when a hypothesis is refuted and no other lead remains, when
the budget is exhausted, or when progress stalls. It returns the best tested change meeting correctness and performance
requirements, or a report of failed trials, with evidence and open questions.

For the \emph{inference-engine optimizations} in
\tabref{tab:analyzer-categories}, candidate changes include
overlapping serialized prefill-to-decode KV transfers with independent
computation; reusing valid KV prefixes, metadata, and pre-allocated buffers; and
tuning draft counts or pruning unused speculative computation while
preserving required state updates. The agent can also make blocking KV control calls asynchronous and
coordinate DP metadata before dispatching model work to Workers.

For \emph{operator optimizations}, our internal kernel agent generates or
specializes implementations using an iterative workflow similar to those of
recent open-source agents~\cite{kernelagent,geak2025,atrex-bench}. It targets
production shapes, data types, layouts, and hardware, using shape frequencies
and the baseline to guide optimization. Comparisons cover complete operators
or fusion groups, including preparation and combination, across relevant
shapes. Promising implementations are integrated into the runtime and
reevaluated under the serving workload.

\para{Expert review and deployment.}
Experts examine the returned code and results for semantic correctness and
applicability to the intended deployment. Negative results can redirect
an unsuccessful investigation. Approved changes enter canary validation. Correctness and serving-SLO
monitoring govern wider rollout, with regressions stopping expansion.

%% file: figs/evaluation-overhead-figure.tex
\begin{figure*}[t]
\centering
\includegraphics[width=\linewidth]{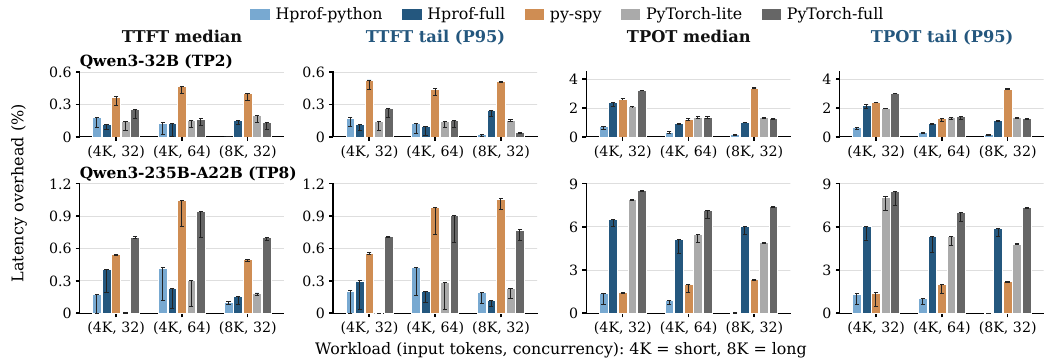}
\caption{Hprof collection overhead on median and tail (P95) latency for
Qwen3-32B~\cite{qwen3-32b-card} and
Qwen3-235B-A22B~\cite{qwen3-235b-card}.
Each bar shows the maximum observed overhead; its whiskers span the observed
range, with non-positive changes plotted as zero.}
\label{fig:eval-cost}
\label{fig:eval-v12-overhead}
\end{figure*}

%% file: section/eval.tex
% Section 8.1 imported from doudun dd-dev 7ca2de1 (2026-09-17).
\section{Evaluation}
\label{sec:eval}

Our evaluation separates collection behavior from the optimization
investigations it supports. We measure Hprof's overhead, lifecycle
behavior, and record fidelity in the tested configurations
(\S\ref{sec:eval-overhead}), then report recurring production findings
(\S\ref{sec:eval-production}). The case studies in
\S\ref{sec:case-studies} show which records supported an analysis,
what additional evidence was needed, and how the resulting constraints
guided a tested modification.

\subsection{Profiling Overhead, Safety, and Fidelity}
\label{sec:eval-profiling}
\label{sec:eval-overhead}
\label{sec:eval-setup}
\para{Testbed and Baseline.}
Each evaluation host has eight NVIDIA H20 GPUs, two 48-core Intel Xeon
Platinum 8469C processors (192 hardware threads), and
1\,TiB of DRAM. The software stack comprises NVIDIA driver 580.105.08,
CUDA 13.0, PyTorch 2.13.0, and vLLM 0.29.0~\cite{vllm}.

We compare Hprof with baselines representing two profiling approaches,
using unprofiled runs as the latency reference:
\begin{enumerate}[label=\arabic*.,leftmargin=1.4em,labelsep=0.4em,
                  itemsep=2.5pt,parsep=0pt,topsep=1.5pt,partopsep=0pt]
  \setlength{\parskip}{0pt}
  \item \emph{Out-of-process stack sampling.}
    We use standalone py-spy~\cite{pyspy} to sample Python call stacks
    and GIL state from outside the target process, at 100\,Hz in blocking mode.
  \item \emph{Framework-integrated profiling.}
    We use PyTorch Profiler~\cite{pytorchprofiler,kineto} to record framework
    operators and CPU/CUDA activities.
    \emph{PyTorch-lite}~\cite{pytorchprofiler} disables shape and stack recording;
    \emph{PyTorch-full}~\cite{pytorchprofiler} enables both.
\end{enumerate}

\para{Workloads.}
We colocate prefill and decode in the same vLLM instance.
The latency experiments cover Qwen3-32B~\cite{qwen3} with TP2 and
Qwen3-235B-A22B~\cite{qwen3} with TP8, both in BF16. Short and long
inputs contain 4,096 (4K) and 8,192 (8K) tokens, respectively;
all requests generate 512 output tokens. Inference concurrency is
32 or 64 for short inputs and 32 for long inputs.

\para{Configurations.}
We evaluate two Hprof configurations. \emph{Hprof-python} records
Python calls and tensor shapes. \emph{Hprof-full} adds
accelerator API calls and GPU activity~(\S\ref{sec:wprof-collection}).

\para{Inference latency overhead.}
Hprof-full has \emph{controllable} overhead: below 0.5\% for TTFT and 7\% for TPOT, as shown in 
Figure~\ref{fig:eval-cost}. With prefill and decode colocated,
the impact is more pronounced on TPOT, particularly for the
larger 235B model. Even there, TPOT overhead is about 20--30\% lower than that of PyTorch-full across the
tested workloads.
Note that these measurements cover active collection; bounded
windows, typically 30\,s, avoid the continuous cost of always-on
tracing (\S\ref{sec:wprof-services}).
Furthermore, we attribute most of the TPOT overhead on large models
such as Qwen3-235B-A22B~\cite{qwen3-235b-card} to CUPTI-based
accelerator tracing~\cite{nvidia-cupti}. Disabling it leaves both
median and P95 TPOT overhead below 1.5\%.
We also noticed that Hprof-python incurs \emph{lower overhead} on
TPOT than py-spy's out-of-process stack sampling, while providing
\emph{richer} records of individual call intervals and tensor shapes.

\para{Resource usage and trace generation.}
Table~\ref{tab:eval-resources} shows Hprof's low CPU usage,
memory footprint, and trace generation rate. Across the tested
workloads, both configurations stay below 0.5 CPU cores,
3\,GiB of peak RSS, and 10\,MiB/s of trace output.
We also consider the total trace volume: at Hprof-full's highest
observed rate, a 30\,s collection window produces approximately
265\,MiB of trace data.
\ifdefined\EvalTextOnly\else
\par\smallskip\noindent\begin{minipage}{\columnwidth}
\centering\small
\setlength{\tabcolsep}{1.2pt}
\captionof{table}{Host footprint and trace production.}
\label{tab:eval-resources}
\begin{tabular*}{\columnwidth}{@{\extracolsep{\fill}}lrrr@{}}
\toprule
Configuration & CPU (cores) & Peak RSS (GiB) & Trace rate (MiB/s)\\
\midrule
Hprof-python & 0.099--0.41 & 2.2--2.8 & 0.36--2.3\\
Hprof-full & 0.12--0.46 & 2.3--2.8 & 1.5--8.8\\
py-spy~\cite{pyspy} & 0.69--1.2 & 2.4--2.7 & 0.01--0.02\\
PyTorch-lite~\cite{pytorchprofiler} & 0.093--0.41 & 3.8--6.1 & 8.5--42\\
PyTorch-full~\cite{pytorchprofiler} & 0.095--0.44 & 5.0--9.3 & 17--87\\
\bottomrule
\end{tabular*}
\vspace{2pt}
\begin{minipage}{\columnwidth}\scriptsize
\end{minipage}
\end{minipage}\par\smallskip
\fi

\para{Profiling Fidelity.}
We compare Hprof with PyTorch Profiler~\cite{pytorchprofiler}
on Qwen3.5-35B-A3B~\cite{qwen35-35b-card}.
Simultaneous collection with Hprof-python and PyTorch Profiler yields 99.2\%
agreement in per-symbol call counts and 99.98\% in immediate-parent
relationships.
For GPU timelines, separate runs of Hprof and PyTorch Profiler on
the same workload yield identical kernel execution order.
Kernel durations span \(0.5\,\mu\mathrm{s}\) to several milliseconds;
absolute differences are approximately 50--150\,ns. Normalizing
these differences by the reference durations gives a median of
1.6\% and a P95 of 10\%.

\para{Lifecycle Safety.}
We evaluated Hprof's lifecycle safety over 50
attach--collect--detach cycles on
Qwen3.5-35B-A3B. Monitoring covered execution during attachment, active collection,
detachment, and subsequent inference. We observed no worker
hangs or inference interruptions during these stages.

% Queue Table 3 for the left column at the top of page 9.
\begin{table}[t]
  \centering
  \footnotesize
  \setlength{\tabcolsep}{3pt}
  \caption{Production optimization outcomes.}
  \label{tab:eval-production-outcomes}
  \begin{tabular}{@{}>{\raggedright\arraybackslash}m{0.28\columnwidth}
                    >{\raggedright\arraybackslash}m{0.30\columnwidth}
                    >{\raggedright\arraybackslash}m{\dimexpr0.42\columnwidth-4\tabcolsep\relax}@{}}
    \toprule
    Optimization & Deployment scope & Measured impact \\
    \midrule
    Overlap KV creation with execution (\S\ref{sec:eval-kvcache})
      & GLM/Qwen KV-backed prefill across $\sim$50 deployments
      & 3.3--5.7\% higher prefill throughput with short contexts \\
    \midrule
    Coordinate DP before Worker dispatch (\S\ref{sec:eval-hybrid})
      & Original Hybrid patch: $\sim$50 deployments
      & Upstream port: 2.5--6.0\% higher token throughput in offline tests \\
    \midrule  
    Prune unused prefill speculation (\S\ref{sec:eval-mtp-prefill})
      & Hybrid-attention models across $\sim$100 deployments (prefill)
      & $\sim$6.0\% TTFT reduction in short-context scenarios  \\
    \midrule
    Optimize Paged Attention data layouts (\S\ref{sec:eval-operators})
      & Small dense models on XPU-C across $\sim$30 deployments
      & $\sim$61\% lower PA duration;
        8--32\% lower~TPOT \\
    \bottomrule
  \end{tabular}
  \vspace{2pt}
  \begin{minipage}{\columnwidth}\scriptsize
  % \textsuperscript{a}September 15 estimates use deployment configurations,
  % image dates, and template-inferred settings; runtime activation was
  % not verified per instance.
  % The two scopes overlap.\\
  % \textsuperscript{b}In offline decode tests on eight XPU-A devices,
  % we take each Qwen model's largest paired-run gain at concurrency 32
  % and at 128, then average the two. The 3.5\% figure gives the three
  % models equal weight.
  \end{minipage}
\end{table}

\subsection{\sys in Production}
\label{sec:eval-production}

Over six months, we collected $\sim$17,000 production traces from over
120 model variants across dozens of clusters and more than 10 accelerator
types, including NVIDIA GPUs and in-house accelerators.

\para{Representative Production Findings.}
Of the $\sim$17,000 collected traces, over 3,900 (23\%)
matched at least one inefficiency pattern.
The collected traces expose issues in host submission, coordination,
and state reuse, alongside operator inefficiencies tied to input shapes
and implementation choices. The following findings describe the serving
conditions in which these issues arise.

\begingroup
\setlength{\parskip}{0pt}
\par\addvspace{3pt}\noindent
\emph{Speculation needs to be judged by useful output.}
In PD-disaggregated hybrid-attention services~\cite{qwen38-max-hf,qwen38-flash-hf,qwen38-flash-blog},
some of the Prefill draft proposals never reach Decode, and extra KV entries extend
beyond the accepted prefix. Pruning must preserve reusable prefix state
(\S\ref{sec:eval-mtp-prefill}).

% \par\addvspace{3pt}\noindent
% \emph{Communication depends on parallelism and data movement.}
% On the PCIe-based accelerators~\cite{rtxpro5000},
% text services use tensor-parallel groups of two or four GPUs, and
% \mbox{MegaMoE} fuses communication with expert computation. Observed
% communication costs are modest. Video generation~\cite{wan-family-hf}
% instead repeatedly gathers weights under FSDP~\cite{pytorch-fsdp}, and
% transfers contend with all-to-all traffic. Reducing traffic or retaining
% gathered weights within a request can help when memory permits.

\par\addvspace{3pt}\noindent
\emph{Submission gaps accompany fragmented execution.}
Dense models below 10B parameters~\cite{qwen3-06b-card,qwen3-4b-card,qwen3-8b-card}
show submission gaps and separate small-kernel launches; Qwen
vision-language profiles show similar gaps~\cite{qwen3-vl-30b-a3b-card}.
Video profiles contain copies, conversions, and layout changes around
communication, motivating fusion and tiled transposes. Broader CUDA Graph
capture can reduce submissions, subject to capture validity and padding costs.

\par\addvspace{3pt}\noindent
\emph{State reuse must account for preparation and capacity.}
Repeated planning, tuning, compilation, and Graph construction offer reuse
opportunities. Attention plans provide reusable metadata for compatible
layers~\cite{flashinfer-plan-reuse}. For KV reuse, savings from cache-aware
routing and shared storage must exceed lookup, transfer, and retained-memory
costs.

\par\addvspace{3pt}\noindent
\emph{Operator implementations must fit online shapes.}
Some attention calls on newer GPUs use older-architecture compatibility
paths; small shapes launch fewer threadblocks than SMs. This motivates
architecture-specific implementations and shape-aware tiling or splitting;
comparisons include conversion and partial-result combination costs.
\par\endgroup

\para{Production Outcomes.}
Guided by these findings, we worked with domain expert teams to evaluate
optimizations under production workloads through canary deployments before
broader rollout. Several are now widely deployed.
\tabref{tab:eval-production-outcomes} summarizes four representative
optimizations, their deployment scope, and measured impact;
\S\ref{sec:case-studies} presents the corresponding analyses and
implementation changes.

% Queue Figure 7 after Table 3 so it occupies the right column at the top of page 9.
\input{figs/kv-supplementary-figure}

%% file: figs/kv-supplementary-figure.tex
\begin{figure}[ht]
  \centering
  \includegraphics[width=\columnwidth]{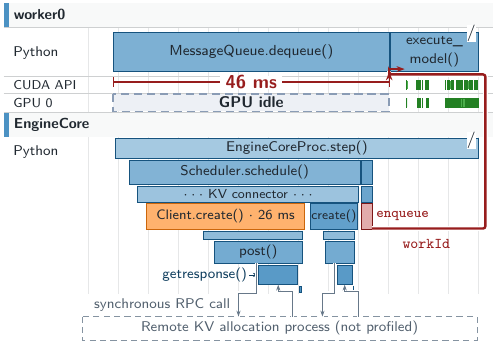}
  \caption{Supplementary profiling links EngineCore and Worker activity.}
  \label{fig:eval-kvcache-trace}
\end{figure}

%% file: section/case-studies.tex
\section{Case Studies}
\label{sec:case-studies}

\label{sec:eval-cases}

The following four case studies trace how runtime evidence supports tested
optimizations: identifying an inefficiency mechanism
(\S\ref{sec:eval-kvcache}), adapting a reference fix to the current execution
constraints (\S\ref{sec:eval-hybrid}), separating unused GPU work from required
state preparation (\S\ref{sec:eval-mtp-prefill}), and optimizing an operator
for production input shapes (\S\ref{sec:eval-operators}).

\subsection{From Profiling to Inefficiency Mechanisms}
\label{sec:eval-kvcache}

We revisit the GLM-5.2~\cite{glm52,glm52-blog} prefill service in
\figref{fig:eval-kvcache-evidence}, where Workers wait for execution
commands while GPUs remain idle. The profiled vLLM instance uses four
of our in-house accelerator devices (XPU-A\footnote{We use XPU as a desensitized name for non-NVIDIA GPUs.}) with TP4.

\para{Initial profiling and observation.}
Hprof's GPU activity records reveal idle intervals in a 25-s
window. Analyzer uses host CUDA API records to associate GPU
execution with the Worker's Python calls, then finds that waits
in \texttt{MessageQueue.dequeue} coincide with 13.8\,s of GPU
idle. Our AI agent inspects the source-code and confirms that this call receives
EngineCore's model-execution commands and that command receipt
gates model execution. Together, these observations support the
\emph{Worker message-dequeue wait} pattern. As
\figref{fig:eval-kvcache-pattern} summarizes, the remaining
question concerns the sender: what delays EngineCore's command?

\para{Supplementary profiling.}
In the next profiling window, we extend Hprof's coverage to EngineCore
scheduling and KV-connector calls. We also collect framework
events carrying \texttt{workId}, which is used to link command
publication to Worker execution. Same-thread call nesting places two synchronous object-creation
calls within the schedule, before command publication, as shown in
\figref{fig:eval-kvcache-trace}. Code inspection explains why this
ordering is inefficient: the objects hold KV after computation and
are not prerequisites for computation itself. EngineCore nevertheless
waits for creation before dispatching, delaying work the Workers
could already execute.

\input{figs/kv-combined-compact-figure}

\para{Fix and impact.}
The reference fix runs KV-hit lookup concurrently with work on other ready
requests~\cite{vllm-pr45659}. We apply the same overlap principle to
object creation: EngineCore starts creation asynchronously and dispatches
model execution while creation proceeds. KV saving still waits for
successful creation, as shown in \figref{fig:eval-kvcache-fix}.
This optimization increases short-context prefill throughput by
3.3\% for GLM-5.2. Across this model and three Qwen
models~\cite{qwen35-397b,qwen3-235b-card,qwen35-35b-card}, the gains
range from 3.3\% to 5.7\%. The change is deployed in these services
on our production MaaS platform.

\input{section/eval-dp-sync-case}
\input{section/eval-mtp-prefill-case}

\input{section/eval-pa-implementation}

%% file: figs/kv-combined-compact-figure.tex
\begin{figure}[t]
  \centering
  \input{figs/kv-combined-compact}
  \caption{Schematic KV investigation and asynchronous creation, with readiness enforced at KV saving.}
  \label{fig:eval-kvcache-combined}
\end{figure}
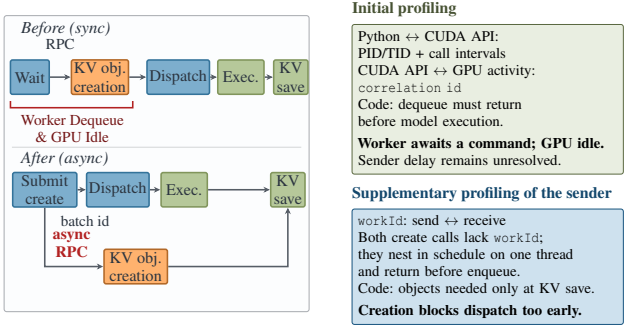

%% file: figs/kv-combined-compact.tex
\begingroup
\input{figs/eval-kvcache-cross-process-styles}
\input{figs/kv-selected-styles}
\subfloat[Asynchronous object creation.\label{fig:eval-kvcache-fix}]{%
\begin{minipage}[t]{.49\linewidth}\centering
\parbox[c][4.70cm][c]{\linewidth}{\centering\resizebox{\linewidth}{!}{\input{figs/kv-fix-selected-original}}}
\end{minipage}}\hfill
\subfloat[Profiling and code evidence.\label{fig:eval-kvcache-pattern}]{%
\begin{minipage}[t]{.49\linewidth}\centering
\parbox[c][4.70cm][c]{\linewidth}{\centering\resizebox{\linewidth}{!}{\input{figs/kv-evidence-compact}}}
\end{minipage}}
\endgroup

%% file: figs/kv-fix-selected-original.tex
% Compact schematic panel; saving retains its object-readiness dependency.
\begin{tikzpicture}[kvcachefigure]
\path[use as bounding box] (0,0) rectangle (4.12,-4.00);
\draw[draw=kvink!40,fill=kvpanelbg,rounded corners=1pt]
  (.015,-.015) rectangle (4.105,-3.985);
\node[kvphase] at (.12,-.20) {Before (sync)};
\node[kvphase] at (.12,-1.94) {After (async)};
\begin{scope}[yshift=-.24cm]
\node[kvhost,minimum width=.52cm,minimum height=.46cm]
  (waitCB) at (.36,-0.6) {Wait};
\node[kvcreate,minimum width=.67cm,minimum height=.46cm]
  (createB) at (1.30,-0.6) {KV obj.\\creation};
\node[kvhost,minimum width=.72cm,minimum height=.46cm]
  (dispatchCB) at (2.35,-0.6) {Dispatch};
\node[kvgpu,minimum width=.62cm,minimum height=.46cm]
  (computeCB) at (3.19,-0.6) {Exec.};
\node[kvgpu,minimum width=.46cm,minimum height=.46cm]
  (saveCB) at (3.86,-0.6) {KV\\save};
\draw[kvflow] (waitCB.east)--node[midway,above=7pt,kvsmall] (syncRpc) {RPC}(createB.west);
\draw[kvflow] (createB.east)--(dispatchCB.west);
\draw[kvflow] (dispatchCB.east)--(computeCB.west);
\draw[kvflow] (computeCB.east)--(saveCB.west);
\draw[kvdelay] (.10,-0.92)--(.10,-1.01)--(1.74,-1.01)--(1.74,-0.92);
\node[kvdelaylabel] at (.93,-1.29)
  {Worker Dequeue\\\& GPU Idle};
\draw[draw=kvink!20] (.12,-1.54)--(4.00,-1.54);
\end{scope}
\begin{scope}[yshift=-.38cm]
\node[kvhost,minimum width=.86cm,minimum height=.46cm]
  (submitC) at (.56,-1.97) {Submit\\create};
\node[kvhost,minimum width=.72cm,minimum height=.46cm]
  (dispatchCA) at (1.54,-1.97) {Dispatch};
\node[kvgpu,minimum width=.62cm,minimum height=.46cm]
  (computeCA) at (2.43,-1.97) {Exec.};
\node[kvgpu,minimum width=.46cm,minimum height=.46cm]
  (saveCA) at (3.82,-1.97) {KV\\save};
\draw[kvflow] (submitC.east)--(dispatchCA.west);
\node[kvsmall] at (1.10,-2.39) {batch id};
\draw[kvflow] (dispatchCA.east)--(computeCA.west);
\draw[kvflow] (computeCA.east)--(saveCA.west);
\node[kvcreate,minimum width=.85cm,minimum height=.46cm]
  (createA) at (1.78,-3) {KV obj.\\creation};
\draw[kvflow] (submitC.south)|-(createA.west);
\node[kvasync] at (.92,-2.7) {async\\RPC};
\draw[kvflow] (createA.east)--(3.82,-3)--(saveCA.south);
\end{scope}
\end{tikzpicture}

%% file: figs/kv-evidence-compact.tex
% Compact rendering of the existing evidence diagram; both rounds retained.
\begingroup
\definecolor{kvrecordgreen}{HTML}{81955C}
\definecolor{kvrecordblue}{HTML}{1F77B4}
\definecolor{kvrecordfill}{HTML}{B6C89A}
\begin{tikzpicture}[x=1cm,y=1cm,
 font=\fontsize{6.0}{6.9}\selectfont,
 heading/.style={anchor=north west,inner sep=0pt,
  font=\fontsize{6.5}{7.3}\selectfont\bfseries},
 box/.style={anchor=north west,rounded corners=1pt,line width=.4pt,
  inner xsep=2.5pt,inner ysep=2.5pt,align=left,text width=105pt}]
\path[use as bounding box] (0,0) rectangle (4.12,-4.70);
\node[heading,text=kvrecordgreen!55!black] at (.05,-.04)
  {Initial profiling};
\node[box,draw=kvrecordgreen!70!black,fill=kvrecordfill!30]
 at (.05,-.36) {Python $\leftrightarrow$ CUDA API:\\
 PID/TID + call intervals\\
 CUDA API $\leftrightarrow$ GPU activity:\\
 \texttt{correlation id}\\
 Code: dequeue must return\\
 before model execution.\\[2pt]
 \textbf{Worker awaits a command; GPU idle.}\\
 Sender delay remains unresolved.};
\node[heading,text=kvrecordblue!60!black] at (.05,-2.67)
  {Supplementary profiling of the sender};
\node[box,draw=kvrecordblue!70!black,fill=kvrecordblue!22]
 at (.05,-2.99) {\texttt{workId}: send $\leftrightarrow$ receive\\
 Both create calls lack \texttt{workId};\\
 they nest in schedule on one thread\\
 and return before enqueue.\\
 Code: objects needed only at KV save.\\[2pt]
 \textbf{Creation blocks dispatch too early.}};
\end{tikzpicture}
\endgroup

%% file: section/eval-dp-sync-case.tex
% !TEX root = ../main.tex
\subsection{Agent-Driven Optimization}
\label{sec:eval-hybrid}

Applying reference fixes still depends on serving conditions. Here the agent adapts synchronization
to Hybrid Attention and prefill/decode (P/D) handoff.

\para{Problem and Mechanism.}
Our MoE deployment uses data parallelism (DP), requiring ranks to coordinate
token counts, padding, and execution modes. A typical implementation uses
Worker-side NCCL, followed by blocking GPU-to-CPU readback. We observe
delayed multi-token prediction (MTP)~\cite{deepseek-v3} submissions, as
shown in \figref{fig:eval-hybrid-before}. As shown in
\figref{fig:eval-hybrid-reference-code}, the matched reference
fix~\cite{vllm-pr29311} adds six lines to select CPU/Gloo instead.
However, Workers still wait for coordination before submitting GPU work.

\para{Agent Adaptation Challenge.}
Moving coordination before Worker dispatch requires EngineCore to
distinguish prefill from decode execution. This is challenging in our
Qwen-family deployment with Hybrid Attention~\cite{qwen35-397b}.
After P/D handoff, the decode instance still processes a few prompt tokens.
The Worker handles this tail through the decode path to continue
transferred recurrent state. An ordinary prefill chunk can
contain just as many tokens, however, so EngineCore's token-count test
cannot distinguish them.

\begin{figure}[t]
\centering
\input{figs/eval-dp-sync-execution-compact-tikz}
\caption{DP coordination before and after adaptation; planning and execution are not time-aligned.}
\label{fig:eval-hybrid}
\end{figure}
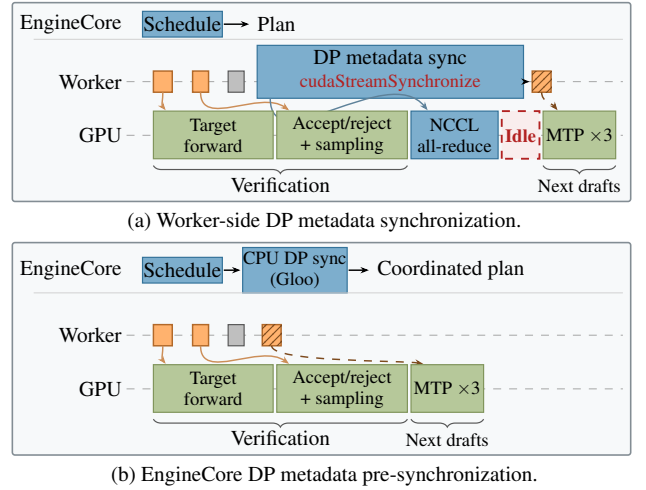

\input{figs/eval-dp-reference-agent-code}

\begin{table}[!b]
\centering\small
\setlength{\tabcolsep}{3pt}
\begin{tabularx}{\columnwidth}{@{}>{\raggedright\arraybackslash}Xrr@{}}
\toprule
Upstream port: context and revision & \shortstack[r]{Runtime\\$+/-$} & \shortstack[r]{Tests\\$+/-$}\\
\midrule
\textbf{Construct the plan.} Pre-dispatch coordination, plan validation, and metadata reuse.
& $688/32$ & $647/1$\\
\midrule
\textbf{Extend communication support.} After inspecting the backend, the agent replaces a blanket expert-parallel
exclusion with a specific applicability check.
& $17/3$ & $22/0$\\
\midrule
\textbf{Repair activation.} Execution logs expose stale width checks;
both producer and consumer use actual draft counts.
& $13/8$ & $23/0$\\
\midrule
Final diff from upstream & $707/32$ & $692/1$\\
\bottomrule
\end{tabularx}
\caption{\DPAgentIcon\hspace{2pt}Agent-driven Optimization Stages.}
\label{tab:dp-feedback}
\end{table}

% \para{Agent-driven Optimization Loop.}
% The agent recovers request boundaries and progress so that EngineCore
% can distinguish a handoff tail from an ordinary prefill chunk with the
% same token count. EngineCore prepares the coordinated execution plan before
% Worker dispatch, as shown in \figref{fig:eval-hybrid-agent-code}.
% The Worker uses this plan for target and draft execution, moving metadata
% coordination out of its target-to-draft submission path
% (\figref{fig:eval-hybrid-after}).
% The Worker retains tail sampling, derives first/later draft plans for
% active and idle ranks, and checks graph and fusion eligibility. This adaptation spans eight runtime files ($+766/-150$
% lines), with activation logs and boundary tests.
% \tabref{tab:dp-feedback} records three versions and two feedback-driven
% extensions of the subsequent upstream adaptation. Backend code
% reveals where communication support can be broadened; execution logs then
% show that stale width checks prevent the plan from taking effect. The
% agent repairs both producer and consumer checks and retests.

\para{Agent-driven Optimization Loop.}
The agent recovers request boundaries and progress so that EngineCore can
distinguish a handoff tail from an ordinary prefill chunk with the same token
count. EngineCore then prepares the coordinated execution plan before Worker
dispatch, as shown in \figref{fig:eval-hybrid-agent-code}. The plan supplies
metadata for target and draft execution, moving coordination out of the
target-to-draft submission path shown in
\figref{fig:eval-hybrid-after}. The Worker retains tail sampling, derives
first/later draft plans for active and idle ranks, and checks graph and fusion
eligibility.
\tabref{tab:dp-feedback} details the subsequent upstream port. Backend
inspection identifies where communication support can be broadened, while
execution logs reveal stale draft-width checks that prevent the plan from
taking effect. The agent updates the producer and consumer checks to use
actual draft counts and retests.

\para{Validation and Deployment.}
The original Hybrid patch is deployed. Offline tests of the subsequent
upstream port on eight XPU-A devices yield throughput gains of
\textbf{6.0\%, 4.8\%, and 2.5\%} for the 35B, 122B, and 397B Qwen models,
respectively. Each value is the mean of two maximum observed paired gains, one at 32
concurrent requests and one at 128.

%% file: figs/eval-dp-sync-execution-compact-tikz.tex
% Version 3 selected by the user on 2026-09-14; full-height originals retained.
\begingroup
\usetikzlibrary{patterns,decorations.pathreplacing}
% Match the soft host/compute palette in Figure 9.
% Hatching retains the target/MTP launch distinction in grayscale.
\definecolor{dpcpu}{HTML}{7DAECD}
\definecolor{dplaunch}{HTML}{FFB26E}
\definecolor{dpmtplaunch}{HTML}{B26929}
\definecolor{dpgpu}{HTML}{B6C89A}
\definecolor{dphostedge}{HTML}{34739A}
\definecolor{dpgpuedge}{HTML}{81955C}
\definecolor{dpwait}{HTML}{B22222}
% Match the flat panel treatment in Figure 6.
\definecolor{dppanelink}{RGB}{30,48,61}
\definecolor{dppanelbg}{RGB}{247,249,250}
\tikzset{dptimeline/.style={
  x=1cm,y=1cm,font=\fontsize{8}{9}\selectfont,
  every node/.style={align=center,inner sep=1pt,outer sep=0pt},
  flow/.style={-{Stealth[length=3.2pt]},line width=.55pt},
  lane/.style={black!35,dashed,line width=.35pt},
  panel/.style={draw=dppanelink!55,fill=dppanelbg,semithick,
    rounded corners=1pt},
  host/.style={draw=dphostedge,fill=dpcpu,line width=.45pt},
  phase/.style={decorate,decoration={brace,mirror,amplitude=3pt},
    draw=black!65,line width=.45pt},
  prepare/.style={draw=black!65,fill=black!25,line width=.45pt},
  dpsync/.style={draw=dphostedge,fill=dpcpu,line width=.45pt},
  vlaunch/.style={draw=dpmtplaunch,fill=dplaunch,line width=.55pt},
  mlaunch/.style={draw=dpmtplaunch,fill=dplaunch,line width=.55pt,
    postaction={pattern=north east lines,pattern color=dpmtplaunch!70!black}},
  gpu/.style={draw=dpgpuedge,fill=dpgpu,line width=.45pt}
}}
\captionsetup[subfloat]{captionskip=2pt,nearskip=0pt,farskip=0pt}
\subfloat[Worker-side DP metadata synchronization.\label{fig:eval-hybrid-before}]{%
\input{figs/eval-dp-sync-before-compact-tikz.tex}}
\par\vspace{1pt}
\subfloat[EngineCore DP metadata pre-synchronization.\label{fig:eval-hybrid-after}]{%
\input{figs/eval-dp-sync-after-compact-tikz.tex}}
\par\vspace{1pt}
\begin{tikzpicture}[dptimeline]
\path[use as bounding box] (0,-.16) rectangle (8.34,.16);
\draw[vlaunch] (0.040,-0.100) rectangle (0.250,0.100);
\node[anchor=west,font=\fontsize{6.5}{7.5}\selectfont,align=left,inner sep=0pt] at (0.360,0.000) {Target/sampling launch};
\draw[prepare] (2.590,-0.100) rectangle (2.800,0.100);
\node[anchor=west,font=\fontsize{6.5}{7.5}\selectfont,align=left,inner sep=0pt] at (2.910,0.000) {Prepare draft inputs};
\draw[dpsync] (4.810,-0.100) rectangle (5.020,0.100);
\node[anchor=west,font=\fontsize{6.5}{7.5}\selectfont,align=left,inner sep=0pt] at (5.130,0.000) {DP metadata sync};
\draw[mlaunch] (6.880,-0.100) rectangle (7.090,0.100);
\node[anchor=west,font=\fontsize{6.5}{7.5}\selectfont,align=left,inner sep=0pt] at (7.200,0.000) {MTP launch};
\end{tikzpicture}
\endgroup

%% file: figs/eval-dp-sync-before-compact-tikz.tex
\begin{tikzpicture}[dptimeline]
\path[use as bounding box] (0,-0.71) rectangle (8.34,1.99);
\draw[panel] (.08,-0.675) rectangle (8.26,1.955);
\begin{scope}[xshift=.17cm,xscale=.96]
\node[anchor=east] at (1.45,1.67) {EngineCore};
\draw[host] (1.70,1.500) rectangle (2.80,1.840);
\node at (2.25,1.67) {Schedule};
\draw[flow] (2.82,1.67)--(3.17,1.67);
\node[anchor=west] at (3.25,1.67) {Plan};
\draw[black!18,line width=.35pt] (.20,1.47)--(8.32,1.47);
\node[anchor=east] at (1.45,.91) {Worker};
\node[anchor=east] at (1.45,.20) {GPU};
\foreach \yy in {.91,.20} \draw[lane] (1.48,\yy)--(8.32,\yy);
\draw[vlaunch] (1.85,.77) rectangle (2.11,1.05);
\draw[flow,draw=dplaunch!80!black] (1.98,.77) to[out=-90,in=130] (2.02,.53);
\draw[vlaunch] (2.39,.77) rectangle (2.61,1.05);
\draw[flow,draw=dplaunch!80!black] (2.50,.77) to[out=-90,in=145] (3.72,.53);
\draw[prepare] (2.88,.77) rectangle (3.10,1.05);
\draw[dpsync] (3.28,.69) rectangle (6.95,1.39);
\node at (5.115,1.21) {DP metadata sync};
\node[text=dpwait,font=\fontsize{7}{8}\selectfont] at (5.115,.90) {cudaStreamSynchronize};
\draw[flow,draw=dpcpu!75!black] (3.46,.69) to[out=-90,in=130] (5.63,.53);
\draw[flow] (6.96,.91)--(7.035,.91);
\draw[mlaunch] (7.07,.77) rectangle (7.33,1.05);
\draw[flow,draw=dpmtplaunch!70!black,dashed] (7.20,.77) to[out=-90,in=130] (7.42,.53);
\draw[gpu] (1.85,-.11) rectangle (3.50,.52);
\node[font=\fontsize{7}{8}\selectfont] at (2.675,.20) {Target\\forward};
\draw[gpu] (3.55,-.11) rectangle (5.35,.52);
\node[font=\fontsize{7}{8}\selectfont] at (4.45,.20) {Accept/reject\\+ sampling};
\draw[dpsync] (5.40,-.11) rectangle (6.60,.52);
\node[font=\fontsize{7}{8}\selectfont] at (6.00,.20) {NCCL\\all-reduce};
\draw[draw=dpwait,dashed,fill=dpwait!8,line width=.7pt] (6.65,-.11) rectangle (7.17,.52);
\node[text=dpwait,font=\fontsize{7}{8}\selectfont\bfseries] at (6.91,.20) {Idle};
\draw[gpu] (7.22,-.11) rectangle (8.22,.52);
\node[font=\fontsize{7}{8}\selectfont] at (7.720000000000001,.20) {MTP $\times 3$};
\draw[phase] (1.85,-.20)--(5.35,-.20);
\node at (3.60,-.45) {Verification};
\draw[phase] (7.22,-.20)--(8.22,-.20);
\node[font=\fontsize{7}{7.5}\selectfont] at (7.720000000000001,-.47) {Next drafts};
\end{scope}
\end{tikzpicture}

%% file: figs/eval-dp-sync-after-compact-tikz.tex
\begin{tikzpicture}[dptimeline]
\path[use as bounding box] (0,-0.71) rectangle (8.34,2.17);
\draw[panel] (.08,-0.675) rectangle (8.26,2.135);
\begin{scope}[xshift=.17cm,xscale=.96]
\node[anchor=east] at (1.45,1.77) {EngineCore};
\draw[host] (1.70,1.600) rectangle (2.80,1.940);
\node at (2.25,1.77) {Schedule};
\draw[flow] (2.82,1.77)--(3.05,1.77);
\draw[dpsync] (3.08,1.470) rectangle (4.52,2.070);
\node[font=\fontsize{7}{8}\selectfont] at (3.80,1.77) {CPU DP sync\\(Gloo)};
\draw[flow] (4.54,1.77)--(4.82,1.77);
\node[anchor=west] at (4.90,1.77) {Coordinated plan};
\draw[black!18,line width=.35pt] (.20,1.47)--(8.32,1.47);
\node[anchor=east] at (1.45,.91) {Worker};
\node[anchor=east] at (1.45,.20) {GPU};
\foreach \yy in {.91,.20} \draw[lane] (1.48,\yy)--(8.32,\yy);
\draw[vlaunch] (1.85,.77) rectangle (2.11,1.05);
\draw[flow,draw=dplaunch!80!black] (1.98,.77) to[out=-90,in=130] (2.02,.53);
\draw[vlaunch] (2.39,.77) rectangle (2.61,1.05);
\draw[flow,draw=dplaunch!80!black] (2.50,.77) to[out=-90,in=145] (3.72,.53);
\draw[prepare] (2.88,.77) rectangle (3.10,1.05);
\draw[mlaunch] (3.35,.77) rectangle (3.61,1.05);
\draw[flow,draw=dpmtplaunch!70!black,dashed]
 (3.48,.77) .. controls (3.48,.55) and (5.27,.58) .. (5.57,.53);
\draw[gpu] (1.85,-.11) rectangle (3.50,.52);
\node[font=\fontsize{7}{8}\selectfont] at (2.675,.20) {Target\\forward};
\draw[gpu] (3.55,-.11) rectangle (5.35,.52);
\node[font=\fontsize{7}{8}\selectfont] at (4.45,.20) {Accept/reject\\+ sampling};
\draw[gpu] (5.4,-.11) rectangle (6.4,.52);
\node[font=\fontsize{7}{8}\selectfont] at (5.9,.20) {MTP $\times 3$};
\draw[phase] (1.85,-.20)--(5.35,-.20);
\node at (3.60,-.45) {Verification};
\draw[phase] (5.4,-.20)--(6.4,-.20);
\node[font=\fontsize{7}{7.5}\selectfont] at (5.9,-.47) {Next drafts};
\end{scope}
\end{tikzpicture}

%% file: figs/eval-dp-reference-agent-code.tex
% Requires \usepackage{listings} in the paper preamble.
\begingroup
\definecolor{DPcodeblue}{RGB}{25,65,180}
\definecolor{DPcodered}{RGB}{180,32,27}
\definecolor{DPcodegray}{RGB}{247,247,247}
\lstset{language=Python,basicstyle=\small\ttfamily,
  keywordstyle=\color{DPcodeblue}\bfseries,
  morekeywords=[2]{EngineCore,Worker},keywordstyle=[2]\color{black}\bfseries,
  commentstyle=\color{black}\bfseries,stringstyle=\color{black},
  emph={disable_nccl_for_dp_synchronization,CPU,Gloo,restore,PD_tail_rule,
    Gloo_collect,Gloo_coordinate,derive_draft_plans,stage_maxima,
    all_ranks_have_requests,fused_if_eligible,FULL},
  emphstyle=\color{DPcodered},
  showstringspaces=false,columns=fullflexible,keepspaces=true,
  frame=single,rulecolor=\color{black!25},backgroundcolor=\color{DPcodegray},
  framesep=3pt,xleftmargin=14pt,framexleftmargin=10pt,
  numbers=left,numbersep=4pt,numberstyle=\fontsize{6}{7}\selectfont\color{black!45},
  aboveskip=2pt,belowskip=3pt,breaklines=true,breakatwhitespace=true,breakindent=1em}
\newcommand{\DPh}[1]{\par\addvspace{3pt}\noindent{\sffamily\bfseries\small #1}\par}
\begin{figure}[t]
\setcounter{subfigure}{0}
\DPh{\figurepanellabel{fig:eval-hybrid-reference-code}(a) Reference fix: CPU backend (+6 lines)}
\begin{lstlisting}[firstnumber=1]
# Added configuration rule
if async_scheduling:
    disable_nccl_for_dp_synchronization = True
# Reused Worker CPU path
meta = zeros(4, DP, on=CPU)
meta[:, rank] = [tokens, padded, ubatch, pad]
all_reduce(meta, group=Gloo, op=SUM)
plan = read_padding_and_ubatch(meta)
launch_model(plan)
\end{lstlisting}
\par\medskip\hrule\smallskip
\DPh{\figurepanellabel{fig:eval-hybrid-agent-code}(b) \DPAgentIcon\hspace{2pt}Agent adaptation: Hybrid (+766/\ensuremath{-}150 lines)}
\begin{lstlisting}[firstnumber=1]
# EngineCore: recover request semantics
local = classify(restore(batch), PD_tail_rule)
# Agree on modes, widths, and limits
meta = Gloo_collect(local)  # 13 fields
plan = agree_target_modes_and_widths(meta)
plan.draft = any(meta.draft) and all(meta.length_ok)
dispatch(batch, plan)
# Worker: keep tail and idle-rank rules
target_or_dummy(plan)
if has_PD_tail(batch): sample_tail_once()
first, later = derive_draft_plans(plan)
# Keep graph eligibility; guard MTP fusion
graphs = draft_graphs(stage_maxima(first, later))
fused = old_eligible and plan.uniform \
    and graphs.later == FULL \
    and all_ranks_have_requests(plan)
draft_if_enabled(first, later, graphs, fused)
\end{lstlisting}
\caption{Reference CPU path and agent adaptation.}
\label{fig:eval-hybrid-optimizer-code}
\end{figure}\endgroup

%% file: section/eval-mtp-prefill-case.tex
\subsection{Inefficiency Beyond GPU Idle Time}
\label{sec:eval-mtp-prefill}

\sys goes beyond idle-time analysis to uncover wasted GPU work
that hurts serving efficiency even on busy GPUs.

\begin{figure}[t]
\centering
\includegraphics[width=\columnwidth]{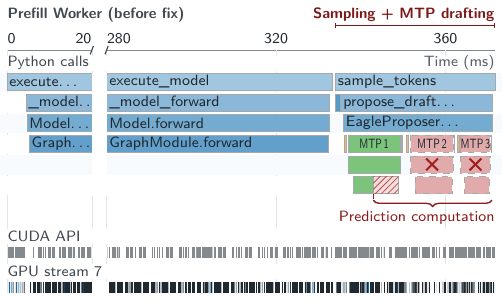}
\caption{Prefill trace: crosses mark removed forwards; hatching marks the prediction path retained within MTP~1.}
\label{fig:eval-mtp-profile}
\end{figure}

\para{Problem and Profiling Observation.}
In our production services for Qwen-family
models~\cite{qwen3-235b-card,qwen35-397b,qwen3-next-80b-card}, we study
a hybrid-attention model combining full attention with Gated DeltaNet
(GDN) layers~\cite{gated-deltanet}. Prefill prepares the target model's
prefix KV and recurrent state for Decode. It also builds the multi-token
prediction (MTP) drafter's own prompt-prefix KV cache, which Decode loads
to resume drafting. Decode generates and verifies candidates locally.
Prefill nevertheless continues proposing tokens after this cache
is ready, keeping the GPU busy with work Decode does not use.
Hprof's Python timings show that sampling and drafting take an average of
39.1\,ms per Prefill-worker step; drafting alone accounts for \textbf{8.8\%} of the combined
model-forward and sampling time. \figref{fig:eval-mtp-profile} shows
GPU work across three draft forwards.

\para{Wasted Computation Despite GPU Activity.}
Our AI agent inspects the code and confirms that Prefill neither transfers speculative candidates to Decode nor verifies them locally.
Later draft forwards---MTP~2--3 in this
profile---therefore extend only a local prediction chain and write cache
entries beyond the prefix accepted by Decode. Once MTP~1 has constructed the
reusable prefix state, these later forwards have no consumer or required side
effect.
In contrast, MTP 1 constructs the transferred prefix state but also produces a prediction unused by Decode. Its
prediction path, marked by hatching in \figref{fig:eval-mtp-profile}, may be
pruned only if every operation needed to construct the transferred state
remains. This intra-forward pruning remains an optimization opportunity.
% We leave this intra-forward pruning as an additional optimization
% opportunity.

\para{Pattern and Optimization.}
The unused candidate-generation path satisfies \emph{Unconsumed output
without required side effects}, a \emph{Speculation / unused work} pattern.
Reference fixes preserve draft-cache updates and suppress subsequent
proposal expansion~\cite{vllm-pr39266,vllm-pr45280,tensorrtllm-pr6104}.
Optimizer's agent similarly removes the two later MTP forwards while retaining the first full forward for state preparation.
This optimization is widely deployed across our
Qwen-family Prefill instances and reduces TTFT by
$\sim$\textbf{6\%} in typical short-context scenarios.

%% file: section/eval-pa-implementation.tex
\subsection{Optimizing Operator Implementations} \label{sec:eval-operators}

At XPU-C's initial release, expert assessment and offline tuning suggested that its
performance was comparable to that of NVIDIA H20. Results for large models~\cite{qwen3-family-hf,qwen25vl72b-card} supported that
expectation, while smaller dense Qwen3 models showed
remaining gaps, as illustrated in \figref{fig:eval-pa-shapes}.

\para{Comparing operator executions.}
We examine PagedAttention (PA) in Qwen3-4B~\cite{qwen3-4b-card}
at a production input shape. Analyzer links Hprof's
Python calls and input fields to host API submissions and GPU activity,
recovering the operator's shape and execution interval. It compares PA's speed-of-light (SOL)
attainment~\cite{solar2026,sol-execbench,roofline2008} on XPU-C with that on
H20.
For the profiled shape, the inputs imply 2.36\,GFLOPs and
0.59\,GB of logical traffic across PA calls per decode token. Using the normalization
in \S\ref{sec:analyzer-evidence}, we obtain 0.76\% MFU and 7.32\% MBU,
hence 7.3\% SOL on XPU-C. H20 exceeds 17\% SOL at the same shape and precision.
The \emph{Low SOL vs.\ reference efficiency} check therefore
directs attention to the operator implementation used for this shape.

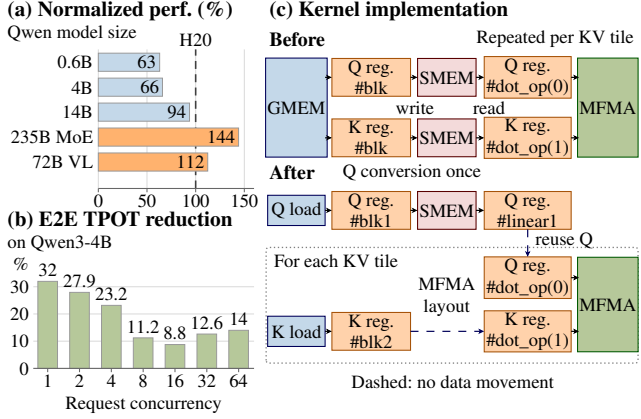
\begin{figure}[t]
\centering
\setcounter{subfigure}{0}
\input{figs/eval-pa-combined-tikz}
\caption{Operator implementation efficiency on XPU-C.}
\label{fig:eval-pa-combined}
\end{figure}

\para{Optimization and impact.} Inspection of the executed implementation reveals
repeated Q/K layout conversions through shared memory in preparation for matrix fused multiply-add
(MFMA) instructions, which multiply matrix tiles and accumulate
results~\cite{amd-mfma-instructions}. Optimizer supplies this diagnosis and the
recovered shape to the Triton kernel agent. The fix in
\figref{fig:eval-pa-layout} converts Q once for reuse and loads K in a compatible
layout, avoiding K's shared-memory round trip~\cite{rtpllm-pa580}.

For the measured shape, this optimization reduces PA operator time by 60.9\% and raises
SOL to 18.7\%, a \textbf{155\% improvement}. \figref{fig:eval-pa-tpot} shows
\textbf{roughly 8--32\% lower end-to-end TPOT} on Qwen3-4B across
request concurrency levels after combining this fix with optimizations to
KV preparation, intermediate copies, and host submission.

%% file: figs/eval-pa-combined-tikz.tex
% Caption/citation revision of the audited Figure 13; native size 8.38 x 5.50 cm.
% All labels >= 7 TeX pt; do not scale down when used in a single column.
% (a) Display labels rounded to integers; bar lengths retain the source values.
%     Values from colleague figs/eval-mi308-operator-optimization-tikz.tex.
%     Selected deployment ratios; configurations are not held constant across models.
% (b) 100 * (release TPOT - optimized TPOT) / release TPOT, paired by concurrency.
%     Entire optimization campaign; not an isolated Attention-kernel benefit.
%     Height uses computed values; labels round to one decimal place.
% Audited source and exact calculations: ../data/eval-pa-metrics.json.
% (c) Original mechanism retained; (B) changed to (c), position translated only.
\begingroup
\definecolor{pasmal}{HTML}{C4D8E9}
\definecolor{paorange}{HTML}{FF7F0E}
\colorlet{palarge}{paorange!60}
\definecolor{pareduce}{HTML}{B6C89A}
\definecolor{waload}{HTML}{C4D8E9}
\colorlet{wareg}{paorange!35}
\definecolor{wasmem}{HTML}{F3DADA}
\definecolor{wamfma}{HTML}{B6C89A}
\begin{tikzpicture}[x=1cm,y=1cm,font=\fontsize{7}{8}\selectfont,
 every node/.style={inner sep=0pt,outer sep=0pt},
 ptitle/.style={font=\fontsize{8}{9}\selectfont\bfseries,anchor=west},
 pgrid/.style={draw=black!15,line width=.3pt},
 paxis/.style={draw=black!60,line width=.4pt}]
\path[use as bounding box] (0,0) rectangle (8.38,5.50);
\node[ptitle] at (.02,5.33) {\figurepanellabel{fig:eval-pa-shapes}(a) Normalized perf. (\%)};
\node[anchor=west] at (.02,5.03) {Qwen model size};

\draw[pgrid] (1.22000,2.98) -- (1.22000,4.82);
\draw[paxis] (1.22000,2.98) -- ++(0,-.05);
\node[anchor=north] at (1.22000,2.9) {0};
\draw[pgrid] (1.86375,2.98) -- (1.86375,4.82);
\draw[paxis] (1.86375,2.98) -- ++(0,-.05);
\node[anchor=north] at (1.86375,2.9) {50};
\draw[pgrid] (2.50750,2.98) -- (2.50750,4.82);
\draw[paxis] (2.50750,2.98) -- ++(0,-.05);
\node[anchor=north] at (2.50750,2.9) {100};
\draw[pgrid] (3.15125,2.98) -- (3.15125,4.82);
\draw[paxis] (3.15125,2.98) -- ++(0,-.05);
\node[anchor=north] at (3.15125,2.9) {150};
\draw[densely dashed,draw=black!75,line width=.6pt] (2.50750,2.98) -- (2.50750,4.82);
\node[anchor=south] at (2.50750,4.86) {H20};
% Qwen3-0.6B: 63.00 percent of H20.
\node[anchor=east] at (1.15,4.65000) {0.6B};
\filldraw[fill=pasmal,draw=black!40,line width=.25pt] (1.22,4.53000) rectangle (2.03112,4.77000);
\node[anchor=east] at (1.98612,4.65000) {63};
% Qwen3-4B: 65.90 percent of H20.
\node[anchor=east] at (1.15,4.32000) {4B};
\filldraw[fill=pasmal,draw=black!40,line width=.25pt] (1.22,4.20000) rectangle (2.06846,4.44000);
\node[anchor=east] at (2.02346,4.32000) {66};
% Qwen3-14B: 93.70 percent of H20.
\node[anchor=east] at (1.15,3.99000) {14B};
\filldraw[fill=pasmal,draw=black!40,line width=.25pt] (1.22,3.87000) rectangle (2.42639,4.11000);
\node[anchor=east] at (2.38139,3.99000) {94};
% Qwen3-235B-A22B: 144.10 percent of H20.
\node[anchor=east] at (1.15,3.66000) {235B MoE};
\filldraw[fill=palarge,draw=black!40,line width=.25pt] (1.22,3.54000) rectangle (3.07529,3.78000);
\node[anchor=east] at (3.03029,3.66000) {144};
% Qwen2.5-VL-72B: 112.40 percent of H20.
\node[anchor=east] at (1.15,3.33000) {72B VL};
\filldraw[fill=palarge,draw=black!40,line width=.25pt] (1.22,3.21000) rectangle (2.66715,3.45000);
\node[anchor=east] at (2.62215,3.33000) {112};
\draw[paxis] (1.22,2.98) -- (3.28,2.98);
\node[ptitle] at (.02,2.56) {\figurepanellabel{fig:eval-pa-tpot}(b) E2E TPOT reduction};
\node[anchor=west] at (.02,2.26) {on Qwen3-4B};
\node[anchor=east] at (.30,1.99) {\%};
\draw[pgrid] (.36,0.60000) -- (3.28,0.60000);
\node[anchor=east] at (.30,0.60000) {0};
\draw[pgrid] (.36,0.96000) -- (3.28,0.96000);
\node[anchor=east] at (.30,0.96000) {10};
\draw[pgrid] (.36,1.32000) -- (3.28,1.32000);
\node[anchor=east] at (.30,1.32000) {20};
\draw[pgrid] (.36,1.68000) -- (3.28,1.68000);
\node[anchor=east] at (.30,1.68000) {30};
% Concurrency 1: 6.91 -> 4.7 ms; reduction 31.982633863965%.
\filldraw[fill=pareduce,draw=black!40,line width=.25pt] (0.42000,0.6) rectangle (0.68000,1.75137);
\node[anchor=south] at (0.55000,1.81637) {32};
\draw[paxis] (0.55000,0.6) -- ++(0,-.05);
\node[anchor=north] at (0.55000,0.52000) {1};
% Concurrency 2: 7.16 -> 5.16 ms; reduction 27.932960893855%.
\filldraw[fill=pareduce,draw=black!40,line width=.25pt] (0.84100,0.6) rectangle (1.10100,1.60559);
\node[anchor=south] at (0.97100,1.67059) {27.9};
\draw[paxis] (0.97100,0.6) -- ++(0,-.05);
\node[anchor=north] at (0.97100,0.52000) {2};
% Concurrency 4: 7.6 -> 5.84 ms; reduction 23.157894736842%.
\filldraw[fill=pareduce,draw=black!40,line width=.25pt] (1.26200,0.6) rectangle (1.52200,1.43368);
\node[anchor=south] at (1.39200,1.49868) {23.2};
\draw[paxis] (1.39200,0.6) -- ++(0,-.05);
\node[anchor=north] at (1.39200,0.52000) {4};
% Concurrency 8: 8.29 -> 7.36 ms; reduction 11.218335343788%.
\filldraw[fill=pareduce,draw=black!40,line width=.25pt] (1.68300,0.6) rectangle (1.94300,1.00386);
\node[anchor=south] at (1.81300,1.06886) {11.2};
\draw[paxis] (1.81300,0.6) -- ++(0,-.05);
\node[anchor=north] at (1.81300,0.52000) {8};
% Concurrency 16: 11.07 -> 10.1 ms; reduction 8.762420957543%.
\filldraw[fill=pareduce,draw=black!40,line width=.25pt] (2.10400,0.6) rectangle (2.36400,0.91545);
\node[anchor=south] at (2.23400,0.98045) {8.8};
\draw[paxis] (2.23400,0.6) -- ++(0,-.05);
\node[anchor=north] at (2.23400,0.52000) {16};
% Concurrency 32: 17.75 -> 15.51 ms; reduction 12.619718309859%.
\filldraw[fill=pareduce,draw=black!40,line width=.25pt] (2.52500,0.6) rectangle (2.78500,1.05431);
\node[anchor=south] at (2.65500,1.11931) {12.6};
\draw[paxis] (2.65500,0.6) -- ++(0,-.05);
\node[anchor=north] at (2.65500,0.52000) {32};
% Concurrency 64: 30.23 -> 26 ms; reduction 13.992722461131%.
\filldraw[fill=pareduce,draw=black!40,line width=.25pt] (2.94600,0.6) rectangle (3.20600,1.10374);
\node[anchor=south] at (3.07600,1.16874) {14};
\draw[paxis] (3.07600,0.6) -- ++(0,-.05);
\node[anchor=north] at (3.07600,0.52000) {64};
\draw[paxis] (.36,.60) -- (3.28,.60);
\node[anchor=south] at (1.82,.015) {Request concurrency};
\begin{scope}[xshift=3.43cm,yshift=.27cm,
 x=1cm,y=1cm,font=\fontsize{7}{8}\selectfont,
 every node/.style={align=center,inner sep=.7pt,outer sep=0pt},
 flow/.style={-{Stealth[length=2.3pt,width=2.1pt]},line width=.45pt},
 cast/.style={flow,dashed,draw=blue!35!black},
 reg/.style={draw=paorange!65!black,fill=wareg,line width=.4pt,minimum width=1.04cm,minimum height=.50cm},
 operand/.style={reg,minimum width=1.15cm},
 load/.style={draw=blue!35!black,fill=waload,line width=.4pt,minimum width=.76cm,minimum height=.40cm},
 smem/.style={draw=red!35!black,fill=wasmem,line width=.4pt,minimum width=.65cm,minimum height=.50cm},
 mfma/.style={draw=green!35!black,fill=wamfma,line width=.4pt,minimum width=.76cm},
 title/.style={font=\fontsize{8}{9}\selectfont\bfseries},
 head/.style={font=\fontsize{7.5}{8.5}\selectfont\bfseries}]

\node[title,anchor=west] at (.03,5.06) {\figurepanellabel{fig:eval-pa-layout}(c) Kernel implementation};
\node[head,anchor=west] at (.03,4.69) {Before};
\node[anchor=east] at (4.90,4.69) {Repeated per KV tile};
\node[load,minimum height=1.31cm] (bg) at (.41,3.78) {GMEM};
\node[reg] (bq) at (1.40,4.18) {Q reg.\\\#blk};
\node[reg] (bk) at (1.40,3.38) {K reg.\\\#blk};
\node[smem] (bsq) at (2.40,4.18) {SMEM};
\node[smem] (bsk) at (2.40,3.38) {SMEM};
\node[operand] (bqo) at (3.47,4.18) {Q reg.\\\#dot\_op(0)};
\node[operand] (bko) at (3.47,3.38) {K reg.\\\#dot\_op(1)};
\node[mfma,minimum height=1.31cm] (bm) at (4.53,3.78) {MFMA};
\draw[flow] (.79,4.18) -- (bq.west);
\draw[flow] (.79,3.38) -- (bk.west);
\draw[flow] (bq) -- (bsq);
\draw[flow] (bk) -- (bsk);
\draw[flow] (bsq) -- (bqo);
\draw[flow] (bsk) -- (bko);
\draw[flow] (bqo.east) -- (4.15,4.18);
\draw[flow] (bko.east) -- (4.15,3.38);
\node at (1.98,3.78) {write};
\node at (2.96,3.78) {read};
\node[head,anchor=west] at (.03,2.92) {After};
\node[anchor=west] at (1.00,2.92) {Q conversion once};
\node[load] (aqg) at (.41,2.42) {Q load};
\node[reg] (aqb) at (1.40,2.42) {Q reg.\\\#blk1};
\node[smem] (aqs) at (2.40,2.42) {SMEM};
\node[operand] (aql) at (3.47,2.42) {Q reg.\\\#linear1};
\draw[flow] (aqg) -- (aqb);
\draw[flow] (aqb) -- (aqs);
\draw[flow] (aqs) -- (aql);
\draw[black!50,densely dotted,line width=.5pt,rounded corners=1pt]
 (.01,.40) rectangle (4.93,1.93);
\node[anchor=west] at (.09,1.75) {For each KV tile};
\node[operand] (aqo) at (3.47,1.53) {Q reg.\\\#dot\_op(0)};
\node[load] (akg) at (.41,.82) {K load};
\node[reg] (akb) at (1.40,.82) {K reg.\\\#blk2};
\node[operand] (ako) at (3.47,.82) {K reg.\\\#dot\_op(1)};
\node[mfma,minimum height=1.25cm] (am) at (4.53,1.175) {MFMA};
\draw[cast] (aql) -- (aqo);
\node[anchor=west] at (3.55,2.02) {reuse Q};
\draw[flow] (akg) -- (akb);
\draw[cast] (akb) -- (ako);
\node at (2.40,1.28) {MFMA\\layout};
\draw[flow] (aqo.east) -- (4.15,1.53);
\draw[flow] (ako.east) -- (4.15,.82);
\node at (2.48,.15) {Dashed: no data movement};

\end{scope}
\end{tikzpicture}
\endgroup

%% file: section/discussion.tex
\section{Experience}
\label{sec:experience}
\label{sec:discussion}
\label{sec:experience-investigations}

\para{Online profiling locates work; deeper analysis explains its cost.}
Online timelines and runtime state expose submission gaps, dependencies that cause waiting, repeated work, and shape-dependent operator costs.
Our program-counter (PC) and performance-monitor (PM) sampling trials
exceeded the online overhead budget, so we reserve instruction-level
analysis for offline profiling with Nsight Compute~\cite{ncu-profiling-guide}.
% Even detailed counters cannot establish whether a proposed change improves
% serving; controlled A/B comparisons with high-overhead profiling disabled must
% test that claim.

\para{Removing detected overhead does not guarantee a performance gain.}
We encountered misdiagnosed bottlenecks and genuine overheads whose proposed fixes yielded no measurable gain.
Experiments must retain the arrival patterns,
shapes, and dependencies that expose the problem; adoption requires repeatable
service gains under correctness and latency constraints. Rejecting one
implementation under one workload does not invalidate the optimization
direction.

% \para{Optimization candidates require end-to-end validation.}
% A measured overhead identifies an optimization opportunity, but removing it
% may not improve serving performance. Experiments must retain the arrival
% patterns, shapes, and dependencies that expose the problem. Only repeatable
% gains under correctness and latency constraints justify adoption.

\para{Successful fixes are fast to review; failures are not.}
% Our agents were most effective at removing explicit
% overheads and adapting known techniques. Pursuing state-of-the-art
% performance still required expert algorithmic and hardware insight.
% Retaining failed attempts and their experimental conditions helps experts
% redirect the search when conditions change.
Our agents were effective at removing explicit overheads and adapting known techniques. 
% Reviewing successful changes is fast—confirmed mechanisms and controlled tests bound the inspection. 
Successful changes were quicker to review because confirmed mechanisms and controlled tests narrowed what experts needed to check.
Failures take longer: the negative result may reflect a wrong diagnosis, a flawed implementation, or masking workload conditions. Retaining failed trials helps experts locate the breakdown.

\para{Profiler-side adaptation to engine evolution.}
Hprof's object-field readers must track changes to engine-internal layouts. 
Updating these readers requires no engine source-code changes, restarts, or profiling support in the engine.
Over six months we released five reader versions to cover
layout changes across engine updates.
% Between reader releases, Hprof still collects call intervals
% and accelerator activity; only shape-dependent observations
% are unavailable until the reader catches up.

%% file: section/related_work.tex
\section{Related Work}
\label{sec:related}

\para{Production profiling.}
GWP~\cite{gwp} established fleet-wide CPU profiling, while Strobelight and
Zoomer~\cite{strobelight,zoomer} support profiling and performance analysis
across production services. Hprof supports inference analysis with runtime attachment
and dynamically configurable, on-demand collection.

\para{Training and inference analysis.}
Training diagnosis systems correlate signals across the execution
stack~\cite{sysomai} or exploit repeated execution to locate performance
problems~\cite{eroica,flare}. For inference, eInfer and
ProfInfer~\cite{einfer,profinfer} provide fine-grained eBPF-based tracing.
% StriaTrace~\cite{striatrace} combines semantic tracing and GPU activity.
\sys complements this analysis with collection whose coverage can expand in
running services as an investigation exposes missing evidence. It then uses
the resulting execution observations to assess inefficiency mechanisms and the
applicability of reference fixes.

\para{Profile-guided and agent-driven optimization.}
Production profiles also guide modifications: AutoFDO~\cite{autofdo} informs
compiler optimizations, and DMon~\cite{dmon} repairs data-locality problems
through selective profiling. ECO~\cite{eco} combines continuous CPU profiles
with mined anti-patterns to develop LLM-generated
optimizations. Code agents~\cite{swe-agent2024} and kernel
agents~\cite{kernelagent,cuda-agent2026,geak2025} automate implementation and
testing. Like ECO, \sys derives suggested solutions from reference fixes.
% current inference evidence establishes their applicability and the dependencies
% its agents must preserve.

%% file: section/conclusion.tex
\section{Conclusion}
\label{sec:conclusion}

We presented \sys, a continuous optimization system for production LLM
inference. Runtime attachment enables rich full-stack profiling on demand
within bounded windows, with coverage refined as investigations progress.
Analyzer reconstructs execution relationships to identify inefficiencies,
and Optimizer guides AI agents through implementation and controlled
validation, with expert review before deployment. Our production
investigations connect execution evidence to engine
and operator changes across diverse serving conditions.